\documentclass[11pt]{article}

\usepackage{amssymb,amsmath,latexsym,eurosym}
\usepackage{amsfonts}
\usepackage[T1]{fontenc}
\usepackage{afterpage}
\usepackage{enumerate}
\usepackage{graphicx}
\usepackage{hyperref}
\usepackage{fancyhdr}
\usepackage{color}
\usepackage{cancel}
\usepackage{graphicx}

\def\reff#1{{\rm(\ref{#1})}}

\def\be{\begin{eqnarray}}
\def\ee{\end{eqnarray}}

\def\b*{\begin{eqnarray*}}
\def\e*{\end{eqnarray*}}

\newtheorem{Theorem}{Theorem}[part]
\newtheorem{Definition}{Definition}[part]
\newtheorem{Proposition}{Proposition}[part]

\newtheorem{Lemma}{Lemma}[part]
\newtheorem{Corollary}{Corollary}[part]
\newtheorem{Remark}{Remark}[part]

\makeatletter \@addtoreset{equation}{section}

\@addtoreset{Definition}{section}

\@addtoreset{Theorem}{section}

\@addtoreset{Proposition}{section}

\@addtoreset{Assumption}{section}

\@addtoreset{Corollary}{section}

\@addtoreset{Lemma}{section}

\@addtoreset{Remark}{section}

\@addtoreset{Example}{section}

\@addtoreset{Exercice}{section}

\makeatother \makeatletter

\def \be{\begin{eqnarray}}
\def \ee{\end{eqnarray}}
\def \b*{\begin{eqnarray*}}
\def \e*{\end{eqnarray*}}
\def\no{\noindent}

\def \E{\mathbb{E}}
\def \F{\mathbb{F}}

\def \L{\mathbb{L}}

\def \N{\mathbb{N}}
\def \P{\mathbb{P}}

\def \R{\mathbb{R}}

\def \[{[\,\!\![}
\def \]{]\,\!\!]}

\def \1{{\bf 1}}

\def \proof{{\noindent \bf Proof. }}
\def \ep{\hbox{ }\hfill$\Box$}

\def\reff#1{{\rm(\ref{#1})}}

\def\Ac{{\cal A}}

\def\Cc{{\cal C}}

\def\Ec{{\cal E}}
\def\Fc{{\cal F}}

\def\Hc{{\cal H}}

\def\Lc{{\cal L}}

\def\Lc{{\cal L}}

\def\Pc{{\cal P}}
\def\Sc{{\cal S}}

\def\eps{\varepsilon}

\def \proof{{\noindent \bf Proof. }}
\def \ep{\hbox{ }\hfill$\Box$}
\def\reff#1{{\rm(\ref{#1})}}

\def\no{\noindent}

\def\={\;=\;}
\def\.{\;.}
\def\be{\begin{eqnarray}}
\def\ee{\end{eqnarray}}
\def\beq{\begin{equation}}
\def\eeq{\end{equation}}

\def\b*{\begin{eqnarray*}}
\def\e*{\end{eqnarray*}}

\def\1{{\bf 1}}

\def\eps{\varepsilon}

\def \E{\mathbb{E}}
\def \F{\mathbb{F}}

\def \N{\mathbb{N}}
\def \R{\mathbb{R}}

\def\P{\mathbb{P}}

\def\Ac{{\cal A}}

\def\Cc{{\cal C}}

\def\Ec{{\cal E}}
\def\Fc{{\cal F}}

\def\Hc{{\cal H}}

\def\Lc{{\cal L}}

\def\Pc{{\cal P}}

\def\Sc{{\cal S}}

\def\eps{\varepsilon}

\def\L{{\cal L}}

\newcommand{\Supp}{ \delta_{K}}
\newcommand{\DomSupp}{ \tilde{K}}
\newcommand{\Lift}[1]{ \textsc{F}_K[#1]}
\newcommand{\set}[1]{ \{#1 \}}
\def \eproof{\hbox{ }\hfill$\Box$}

\newcommand{\HYP}[1]
    {\ensuremath{({\bold H#1} ) }}
\newcommand{\esp}[1]
    {\ensuremath{%
     \mathbb{E}\!\!\left[#1\right]}}

\definecolor{darkred}{rgb}{0.8,0,0}
\definecolor{darkblue}{rgb}{0,0,0.7}
\definecolor{darkgreen}{rgb}{0,0.4,0}

\title{ 
When terminal facelift enforces Delta constraints
\footnote{Research supported by the ANR grant LIQUIRISK and the Chair \textit{Finance and Sustainable Development.}}}
\author{Jean-Fran\c{c}ois Chassagneux\footnote{Department of Mathematics, Imperial College London.  { \sf j.chassagneux@imperial.ac.uk}} \and Romuald Elie\footnote{CEREMADE, Universit{\'e} Paris-Dauphine.{ \sf \{elie,kharroubi\}@ceremade.dauphine.fr}} \and Idris Kharroubi\footnotemark[\value{footnote}]}

\date{This version: July 2013, first version: August 2012}
\begin{document}
\maketitle 
\begin{abstract}

 This paper deals with the super-replication of non path-dependent European claims under additional convex constraints on the number of shares held in the portfolio. 
  The corresponding super-replication price of a given claim has been widely studied in the literature and its terminal value, which dominates the claim of interest, is the so-called facelift transform of the claim. We investigate under which conditions the super-replication price and strategy of a large class of claims coincide with the exact replication price and strategy of the facelift transform of this claim. In dimension $1$, we observe that this property is satisfied for any local volatility model.  In any dimension, we exhibit an analytical necessary and sufficient condition for this property, which combines the dynamics of the stock together with the characteristics of the closed convex set of constraints. To obtain this condition, we introduce the notion of first order viability property for linear parabolic PDEs. We investigate in details several practical cases of interest: multidimensional Black Scholes model, non-tradable assets or short selling restrictions.
\end{abstract}

\vspace{5mm}


\vspace{5mm}

\noindent {\bf MSC Classification (2010):}    93E20, 91G20,  60H30 

%
%

\newpage
\section{Introduction}

 In a complete financial market, the absence of arbitrage opportunities leads to the definition of a unique fair price for any contingent claim using replication arguments. 
This uniqueness property disappears as soon as constraints are introduced in the replication procedure,  see e.g$.$  \cite{karkou96, chesou05} and references therein.  This implies the existence of a closed interval of arbitrage-free prices. A commonly considered prudential pricing methodology consists in selecting the upper bound of this interval.  This so-called super-replication price coincides with the minimal initial amount of money required to constitute an admissible portfolio strategy satisfying the constraints and whose terminal value dominates the claim of interest. The super-replication price under  convex delta constraints has been thoroughly studied in the literature. In \cite{cvikar93}, the authors obtain a dual representation of the super-replication price in terms of a well chosen set of risk neutral probabilities.  In \cite{cvikarson98}, closely related to the previous work, the super-replication price process is shown to be the unique solution of a Backward Stochastic Differential Equation (BSDE) with constraints on the gain process $Z$. All these works mainly rely on probabilistic and duality arguments.  In a Markovian setting, the super-replication price is characterised using direct dynamic programming arguments  and PDE techniques, see \cite{sontou03, bentoumen05}.

In  \cite{brocvison98}, the authors observe that, for the classical Black Scholes model,  the super-replication price of a claim under convex delta constraints coincides with the unconstrained replication price of a so-called facelift transform of this claim. They consider constraints in terms of number of shares in dimension 1, wealth proportion or money amount in any dimension, and exhibit the three corresponding facelifting procedures. In more general Markovian models, the super-replication price function under convex portfolio constraints of a non path-dependent claim interprets as the smallest function above the unconstrained price of the claim, which is stable under the corresponding facelift transform, see  e.g$.$ \cite{bentoumen05}.  The goal of this paper is to state a necessary and sufficient condition under which the noteworthy result of \cite{brocvison98} extends to general local volatility models in dimension $d$.

To exhibit this condition presented in Theorem \ref{maintheorem} below,  we rely on a BSDE representation of the replicating strategy.  We show in Proposition \ref{pr representation}  that the replicating strategy  is the unique solution of a multidimensional linear BSDE with terminal value $\nabla h(X_T)$, where $h$ is a smooth payoff function and $X$ denotes the assets price process.\\
If $h$ satisfies the portfolio constraint, i.e. $\nabla h$ is valued in some convex set $K$, the condition given in Theorem \ref{maintheorem} ensures that the solution of a multi-dimensional linear BSDE with terminal value $\nabla h(X_T)$ is valued in a convex set $K$. Namely, the super-replication price of the claim with payoff $h(X_T)$ under convex delta constraints coincides with its unconstrained replication price. It is crucial to observe that we cannot rely on
classical viability results for BSDEs \cite{bqr00}  since the class of possible terminal value for the BSDE is restricted here to gradient type terminal conditions of the form $\nabla h(X_T)$. The condition obtained in \cite{bqr00} is thus only a sufficient condition for our problem.
Contrary to this paper, we take advantage here of the linear structure of the problem. It appears that the study of the viability condition for the convex set $K$ boils down to the study of the viability condition for the tangent half-spaces to $K$. This makes the proof - in some sense - simpler. Our approach allows us also to remark that, under the exhibited condition of interest,  the super-replication price of an American option with exercise payoff $h(X)$ under convex delta constraints also coincides with its unconstrained replication price.\\
In practice, the payoff function $h$ does not satisfy that $\nabla h$ is valued in $K$ nor any smoothness property. Nevertheless, our main result still holds using the facelift transform of $h$. The proofs in the general case are based on regularisation techniques.

We also discuss in this paper various practical cases which are of interest in Finance, see Section 3. We first observe that the result of \cite{brocvison98} in the Black Scholes model does not extend to the consideration of a financial market with $d>1$ assets. The hypercubes are the only convex set of constraints for which facelifting the payoff allows to get rid of the portfolio constraints in a multidimensional Black Scholes model. This property extends also to most of the common local volatility models, in which each asset follows its own dynamics. In particular, hypercubes include the consideration of non-tradable assets or no short sell restrictions. More specifically, we observe that the only model dynamics in which no short sell restrictions on Asset $1$ can be relaxed using a facelifting procedure are the one for which the quadratic covariations between the other assets do not depend on Asset $1$.

The rest of this paper is organized as follows. In Section \ref{SectionProblem}, we specify the problem formulation and exhibit the main properties of the super-replication methodology and related facelift operator. In Section \ref{SectionMain}, we produce the main result of the paper which gives a tractable analytical necessary and sufficient condition, ensuring that  the exact replication property holds for a large class of payoff functions. We describe practical examples of interest and provide a simple probability-change argument in dimension $d=1$. Focusing on regular payoff functions $h$ stable under the facelifting procedure, Section \ref{SectionViability} is dedicated to the proof of the necessary and sufficient condition for the so-called first order viability property. Namely, the first order viability property ensures that the gradient of the solution of a linear PDE lies in a convex set $K$ on $[0,T]$ as soon as  it does at time $T$. This section revisits arguments of \cite{bqr00}  in our framework. Section \ref{SectionRegu} provides the proof of the main theorem and details in particular the corresponding regularization procedure. The Appendix collects useful properties of the facelift transform and some technical proofs.

\no {\bf Notations.} Any element $x \in \R^d$  will be
identified to a column vector with $i$-th component $x^i$ and
Euclidian norm $|x|$, $(e_i)_{1\le i \le d}$ is  the canonical basis of $\R^d$. We denote by $M_d$ the set of matrices with $d$ lines and $d$ columns, and $S_d$ the subset of symetric elements of $M_d$. For a matrix $A\in M_d$, $\text{Tr}[A]$ denotes its trace, $A^{.j}$ its $j$-th column, $A^{i.}$ its $i$-th row,  and $A^{ij}$ the $i$-th term of  $A^{.j}$.
The transpose of a matrix or a vector $y$ will be denoted $y^\top$. For a vector $x$, $Diag(x)$ denotes the diagonal square matrix with diagonal terms given respectively by $(x^i)_{1\le i \le d}$. For a function $h$ from $\R^d$ to $\R$, 
 we denote by $\partial_xh$  and $\partial^2_{xx}h$  the $d$-dimensional row vector  and the matrix in $M_d$ defined by
\b*
\partial_x h(x) \; = \; \Big(\frac{\partial h(x)}{\partial x_1},\ldots, \frac{\partial h(x)}{\partial x_d}\Big) \qquad &\mbox{and}&\qquad 
\partial^2_{xx}h(x) \; = \; {\Big[\frac{\partial^2 h(x)}{\partial x_i\partial x_j}  \Big]}_{1\leq i,j\leq d}\;,\quad x\in \R^d\;.
\e*
We shall also note $\partial_i h(x) := \frac{\partial h(x)}{\partial x_i}$ , for $1\le i \le d$ and $\nabla h (x)=[\partial_x h(x)]^\top$, $x\in\R^d$. 
$C^1_b(\R^d,\R^q)$ denotes the set of function from $\R^d$ to $\R^q$  which are differentiable with continuous and bounded first derivatives.
 We denote by $\lambda$ the Lebesgue-measure and by $\Ec_.(U)$ the Doleans-Dade exponential of a process $U$.\\
Finally, for a given closed convex set $K$, $d_K$ is the (non-negative) distance function to this set, namely, $d_K:x\in\R^d\mapsto\inf\{|x-y|~:~y\in K\}$.

%
%

\section{Super-replication and facelift properties}\label{SectionProblem}

In this section, we introduce the market model and formulate the super-replication problem under {\it Delta} constraints, namely when the number of shares constituting the portfolio  must remain in a closed convex set.

\subsection{The market model}

We consider a financial market defined on a probability space $(\Omega,\Fc,\P)$,  endowed with a $d$-dimensional brownian motion $W$. 
For $0\leq t\leq T$, we denote   by $\F^t={(\Fc_s^t)}_{t\leq s\leq T}$ the completion of the filtration generated by $(W_{s}-W_t)_{s\geq t}$ and by $\mathcal{P}^t$ the $\sigma$-field of progressively measurable processes associated to $\F^t$. In the sequel, we interpret the probability $\P$ as a pricing measure.

We suppose that the financial market is composed of $d$ risky assets and one non-risky asset, whose interest rate  is assumed to be $0$ for ease of presentation. Up to considering discounted processes, all the results of the paper extend straightaway to financial markets with \emph{deterministic} interest rates.

For an initial condition $(t,x)\in[0,T]\times\R^d$, where $x$ represents the vector value of the $d$ assets at time $t$, the vector of risky asset price process is described by the diffusion  ${(X^{t,x}_s)}_{s\in[t,T]}$ defined as the unique solution of the stochastic differential equation
\be\label{dynamicsRN}
X_s^{t,x} & = & x+\int_t^s \sigma(X^{t,x}_r)dW_r\;,\qquad s\in[t,T]\;,
\ee
where $\sigma:\R^d\rightarrow M_d$ is the volatility function. The Dynkin second order linear differential operator associated to the dynamics \reff{dynamicsRN}, denoted by $\Lc_\sigma$, 
 is given by
\b*
\Lc_\sigma \varphi (t,x) & := & \partial_t\varphi(t,x)+ 
{1\over 2}\text{Tr}\Big(\partial^2_{xx}\varphi(t,x)\sigma(x) \sigma^\top(x)\Big) \;,\quad (t,x)\in[0,T]\times\R^d\;,
\e*
for any $\varphi\in C^{1,2}([0,T]\times\R^d)$. 

\no {We denote by $\textrm{Supp}(\sigma)$ the interior of the the support of the function $\sigma$ \textit{i.e.} the open subset of $\R^d$ defined by
\b*
\textrm{Supp}(\sigma) & := & \big\{ x\in \R^d~:~\sigma(x) \mbox{ is invertible } \big\}\;.
\e* }
\no Throughout this paper, we  work under the condition that the function $\sigma$ is  $C^1_b(\R^d,M_d)$ and shall sometimes use the following assumption:
 
\vspace{2mm} 
 
\no {\noindent \HYP{\sigma} The function $\sigma^{-1}$ is continuous on  $\textrm{Supp}(\sigma)$ and for any $(t,x)\in \textrm[0,T]\times \textrm{Supp}(\sigma)$, the process $\big(\sigma^{-1}(X_s^{t,x})\big)_{s\in[t,T]}$ is well defined.}

\vspace{2mm}

\vspace{2mm}

Starting with an initial wealth $y \ge 0$ at time $t\in[0,T]$, an investment strategy is described by a $\Pc^t$-measurable process $\Delta=(\Delta^1,\ldots,\Delta^d)^\top$ valued in  $\R^d$, where $\Delta^i_s$ represents the number of shares of  asset $i$ detained at time $s\in[t,T]$. 
We denote by $\Ac_{t,x}$ the set of self financing strategies $\Delta$ such that 
\b*
\int_t^T|\sigma(X_r^{t,x})\Delta_r|^2dr & < & \infty\;, \quad \P-a.s.
\e*
The portfolio process corresponding to an initial wealth $y$ at time $t\in[0,T]$ and a self-financing strategy $\Delta\in\Ac_{t,x}$ is denoted ${(Y^{t,x,y,\Delta}_s)}_{s\in[t,T]}$ and satisfies
\b*
Y^{t,x,y,\Delta}_s & = &
 y+\int_t^s\Delta_r^\top dX^{t,x}_r\;, \qquad t\le s \le T \;.
\e*

The set $\Ac_{t,x}^b$ of admissible strategies is given by the strategies $\Delta$ in $\Ac_{t,x}^b$ such that 
the portfolio value $Y^{t,x,y,\Delta}$ is bounded from below by a constant.

\subsection{Super-replication under constraints}

Due to regulatory or structural reasons, we suppose  that the possible investment strategies are restricted to take their values in a deterministic closed convex subset $K$ of $\R^d$ s$.$t$.$ $0 \in K$. 

For any $(t,x)\in[0,T]\times\R^d$, the subset of admissible  constrained strategies $\Ac_{t,x}^K$ is then defined by
\b*
\Ac_{t,x}^K & := & \big\{\Delta \in\Ac^b_{t,x}~:~
d_K(\Delta) = 0 \;,\; \P\otimes \lambda -a.e.
 \big\} \;.
\e*
Observe that the constraint is not imposed on the portfolio value but on the investment strategy itself.

The addition of constraints on the investment strategy implies that exact replication of   a given contingent claim is not always possible, see  e.g$.$  \cite{karkou96}. Here, we intend to focus on super-replication strategies.
\begin{Definition}[Super-replication valuation] \label{de super-replication}
For a measurable function $h:~\R^d\rightarrow\R$ bounded from below, we define the super-replication price of the contingent claim   $h(X_T^{t,x})$ under $K$ constraint at time $t$ by
\b*
v_K^h(t,x) & = & \inf  \left\{ y\in\R~:~\exists \Delta\in \Ac_{t,x}^K \;,~y+ \!\!\int_t^T\!\!\Delta_s^\top dX^{t,x}_s~\geq ~h(X^{t,x}_T)~\P-a.s.\right\}\,,
\e*

for any $(t,x)\in[0,T]\times\R^d$.
\end{Definition}

The super-replication price of a contingent claim has been widely studied in the literature see e.g. \cite{cvikar93, sontou03, bentoumen05}. In our context, a complete characterization of the super-replication price under constraint is given in \cite{bou10} and we will use a supersolution property of $v_K^h$ proved therein. 
\vspace{1mm}
Let us recall that  the support function $\delta_K$  of the convex set $K$ is defined by
\b*
\delta_K(x) & := & \sup_{y\in K} x^\top y  \;, \qquad x\in \R^d\;,
\e* 
whose domain is denoted
$
\DomSupp := \set{ \zeta \in \R^d \;|\; \Supp(\zeta) < +\infty}.
$

\noindent Using the support function of $K$, we define the following global differential operator related to the constraints:
\begin{align*}
\mathcal{C}_K(p) := \inf_{|\zeta|=1, \zeta \in \DomSupp} \Supp(\zeta) - \zeta^\top p  \;.
\end{align*}

\noindent Let us also  introduce 
\begin{Definition}[Facelift]\label{defFacelift}
The facelift operator $\textsc{F}_K$ for the admissibility set $K$ maps  any measurable function $h:\R^d\rightarrow\R$, to its   facelift transform $\textsc{F}_K [h]$, defined by
\b*
\textsc{F}_K [h] (x) & := & \sup_{y\in\R^d} h(x+y)-\delta_K(y) =   \sup_{y\in\tilde{K}} h(x+y)-\delta_K(y)  \;, \qquad x\in\R^d\;.
\e*
\end{Definition} 
We collect in the Appendix, Section \ref{subse facelift}, some useful properties of the Facelift transform.

\noindent In the sequel, we shall use the following assumption related to the payoff function $h$ and its facelift transform:

\noindent \HYP{h} The function $h$ is lower semi-continuous, bounded from below and such that
 \begin{align*}
 \esp{|\Lift{h}(X_T^{t,x})|^2} < \infty \;,\qquad  (t,x) \in [0,T]\times \R^d.
 \end{align*}

\begin{Remark}
{Assumption \HYP{h} is satisfied, e.g., in the following cases.
\\
(i) When $K=[0,\infty)$ i.e. the no short sell constraint and $h$ is the payoff of a Put option. Indeed, $\Lift{h}$ is then  a constant function equal to the strike of the Put option. 
\\
(ii) When $K$ is bounded or $h$ is Lipschitz continuous, since then $F_K[h]$ is Lipschitz-continuous. }
\end{Remark}

\vspace{2mm}

\noindent Let us now recall the following result proved in \cite{bou10}.
\begin{Proposition} \label{pr bruno}
The super-replication price under $K$-constraint $v_K^h(t,x)$ is a viscosity supersolution of the following PDE 
\b*
\min \set{\, - \mathcal{L}_\sigma u(t,x)\,,\, \mathcal{C}_K(\nabla u(t,x)) \; } & = &  0 \;,\quad  (t, x) \in [0,T)\times \R^d
\e*
and satisfies
\b*
v_K^h(T-,x)  & \ge &  \Lift{h}(x)\;,\quad  x \in \R^d\;,
\e*
provided that $h$ is l.s.c, with linear growth and bounded from below.
\end{Proposition}

We conclude this section with a  consequence of the previous result.

\begin{Corollary}\label{co carac utile}
Assume that h is l.s.c, with linear growth  and bounded from below, then
\b*
v_K^h(t,x) & \ge & \esp{\Lift{h}(X_T^{t,x})}\;, \qquad (t,x)\in[0,T)\times \R^d\;.
\e*
\end{Corollary}

\noindent For sake of completeness, we provide a proof in the Appendix, see Section \ref{Appendix2}.

%
%

\section{Relaxing portfolio constraints via terminal facelift} \label{SectionMain}

In this section, we investigate under which conditions, super-hedging any claim under {\it Delta} constraints is equivalent to simply hedge the facelift transform of this claim. We first consider the particular case where the  number of shares for each asset is constrained to stay in-between two constant bounds. In this context, we show that the `replication property' is always satisfied in the one-dimensional case but not systematically in the multi-dimensional case. This motivates the second part of this section which presents in Theorem \ref{maintheorem} a tractable analytical necessary and sufficient condition ensuring this property to hold for general multi-dimensional convex constraints. We finally focus on several practical examples of importance (Black Scholes model, short selling, non-tradable asset,  etc.) in order to emphasize the range of applications for Theorem \ref{maintheorem}, which is the main result of the paper.

\subsection{A motivating example}

 We consider first a simple practical example where the investor faces constant restrictions on the number of shares of each asset held in the portfolio. More precisely, the admissibility set is a closed hypercube given by 
 \begin{align}\label{eq de Kc}
 K_c &:= \Pi_{i=1}^d \;\; [-d_i, u_i] \;, \qquad \mbox{where $(d_i,u_i)\in[0,\infty]^2$,$\;$ for $1\le i \le d$}\;.  
 \end{align}
 Observe that this form of convex set allows to consider, for example, the realistic practical case where short-selling one or several assets is not allowed. It also covers the natural case where some of the assets cannot be traded on the financial market ($d_i=u_i=0$).

 We first focus on the particular case where only one asset is traded ($d=1$). As detailed in the next proposition, a direct probability change argument shows that super-replicating a claim under $K_c$-portfolio constraints simply consists in replicating without constraint the facelift transform of this claim. For sake of simplicity, we consider here payoff functions with  regular facelift transform, but this strong regularity property is relieved in the following subsection, see Theorem \ref{maintheorem}. 
 
 \begin{Proposition}\label{Propo31} Let $d=1$ and $h$ be a payoff function such that  $F_{K_c}[h]\in C^1_b(\R,\R)$. Then, for any starting point $(t,x)\in[0,T]\times\R$,  the super-replicating price and hedging strategy under $K_c$-constraints of the claim $h(X^{t,x}_T)$ coincides with the exact replicating price and unconstrained hedging strategy  of $F_{K_c}[h](X^{t,x}_T)$. 
 \end{Proposition}
 
 \proof 
 Let $(t,x)\in[0,T]\times\R$ and consider a payoff function $h$ such that $F^{K_c}[h]\in C^1(\R,\R)$. By construction of the facelift transform, $\nabla F_{K_c}[h]$ is necessarily valued in $K_c$. We now consider the exact replicating strategy $\Delta^{F_{K_c}[h]}$ of $F_{K_c}[h](X^{t,x}_T)$ and intend to prove that $\Delta^{F_{K_c}[h]}$ is valued in $K_c$ on $[t,T]$. Due to the regularity of $F_{K_c}[h]$, observe that the exact replicating strategy rewrites
 \b*
 \Delta^{F_{K_c}[h]}_s &=& \E \left[ \nabla F_{K_c}[h](X^{t,x}_T) \frac{\nabla X^{t,x}_T}{\nabla X^{t,x}_s} \;\;\Big|\;\;\Fc^t_s \right] \;, \qquad t\le s \le T\;, 
 \e*
 where $\nabla X^{t,x}$ denotes the tangent process of $X^{t,x}$ and satisfies
\b*
 \nabla X^{t,x}_s  & = & 1+ \int_t^s  \nabla \sigma(X^{t,x}_r) \nabla X^{t,x}_r  dW_r \;, \qquad t\le s \le T\;.
\e*
 Since $\sigma$ has bounded derivatives, we deduce that $(\nabla X_s^{t,x})_{t\le s\le T}$ is a positive martingale 
 with constant expectation equal to $1$. Therefore, it also interprets as a probability change on $(\Omega,\Fc^t_T)$ and we denote by $\P^{\nabla X}$ the probability defined by $\frac{d\P^{\nabla X}}{d\P}|_{\Fc^t_T}=\nabla X^{t,x}_T$. This allows us to rewrite directly
 \b*
 \Delta^{F_{K_c}[h]}_s &=& \E^{\P^{\nabla X}} \left[ \nabla F_{K_c}[h](X^{t,x}_T) \;\;\big|\;\;\Fc^t_s \right] \;, \qquad t\le s \le T\;. 
 \e*
Since $\nabla F_{K_c}[h]$ is valued in the convex set $K_c$,  $\Delta^{F_{K_c}[h]}$ is also  valued in $K_c$.
 
  The  hedging strategy of $F_{K_c}[h](X^{t,x}_T)$ being valued in $K_c$, it coincides with the super-hedging strategy under $K_c$-constraints of $h(X^{t,x}_T)$, see Corollary \ref{co carac utile}. Hence,  the super-replicating price of $h(X^{t,x}_T)$ and the replicating price of $F_{K_c}[h](X^{t,x}_T)$ also coincide.  \ep

{
 \begin{Remark} {\rm  Interpreting the gradient of the stock process as a probability change has  already been used for example in \cite{karjea98}. 
}
 \end{Remark}
 }
 
 We now turn to the multi-dimensional case. As detailed in the next proposition, the previous result easily extends to the particular case where each asset has its own dynamics, since the previous arguments can be applied on each component of the price process $X$. 

 \begin{Proposition}\label{Prop_2}  Let $h$ be a payoff function such that  $F^{K_c}[h]\in C^1_b(\R^d,\R)$. Fix $(t,x)\in[0,T]\times\R^d$ and suppose that the dynamics of each asset $X^{t,x,i}$ is given by 
 \b*
 X^{t,x,i}_s &=& x^i + \int_t^s \sigma_i(X^{t,x,i}_r) dW_r \;, \qquad 1\le i\le d\;, \qquad t\le s \le T \;.
 \e*
 Then, the super-replicating price and hedging strategy under $K_c$-constraints of the claim $h(X^{t,x}_T)$ coincides with the replicating price and hedging strategy of $F_{K_c}[h](X^{t,x}_T)$. 
  \end{Proposition}

 \proof Following the same reasoning as in the one-dimensional case, we only require to verify that the exact replicating strategy $\Delta^{F_{K_c}[h]}$ of $F_{K_c}[h](X^{t,x}_T)$ is valued in $K_c$. As in the one dimensional case, we have
 \b*
 \left(\Delta^{F_{K_c}[h]}_s \right)^i &=& \E \left[ \left(\nabla F_{K_c}[h](X^{t,x}_T)\right)^i \frac{\nabla X^{t,x,i}_T}{\nabla X^{t,x,i}_s} \;\;\Big|\;\;\Fc^t_s \right] \;, \qquad t\le s \le T\;,\quad 1 \le i \le d\;,
 \e*
 where $\nabla X^{t,x,i}$ is the tangent process of $X^{t,x,i}$. Due to the particular form of the stock dynamics, each tangent process $\nabla X^{t,x,i}$ is a positive martingale starting from $1$. Observe that, contrary to the one-dimensional proof, we cannot use a common probability change for all the $d$ components of the hedging strategy $\Delta^{F_{K_c}[h]}$. Nevertheless, due to the special form of $K_c$ and the fact that $\nabla F_{K_c}[h](X{t,x}_T)\in K_c$, we work separately on each component. We  compute
 \b*
 -d_i \;=\; \E\left[ -d_i \frac{\nabla X^{t,x,i}_T}{\nabla X^{t,x,i}_s} \;\;\Big|\;\;\Fc^t_s\right] 
 \;\le\;  \left(\Delta^{F_{K_c}[h]}_s \right)^i
 \;\le\; \E\left[ u_i \frac{\nabla X^{t,x,i}_T}{\nabla X^{t,x,i}_s}\;\;\Big|\;\;\Fc^t_s \right] \;=\; u_i \;,
 \e*
 for $t\le s \le T$ and $1 \le i \le d$. Hence, the replicating strategy $\Delta^{F_{K_c}[h]}$ is valued in $K_c$ and the proof is complete.
 \ep
 
 Unfortunately, this nice property does not remain valid for general multi-dimensional stock dynamics. Consider for example the 2-dimensional case where the dynamics of the first asset $X^1$ is given by
 \b*
 d X^1_t &=&  \left(|X^2_t| \wedge \bar\sigma\right) X^1_t dW_t \;,  \qquad  \mbox{ with $\bar\sigma>0$}\;,
 \e*
 and the second asset (the stochastic volatility) is not tradable, i.e. $K_c=\R\times\{0\}$. In this framework, the super-replicating price of a call (or any convex payoff) option on $X^1_T$  is simply the $\bar\sigma$-volatility Black Scholes price of this call, see  e.g$.$ \cite{cviphatou99}. 
 
 Hence, even for hypercube type constraints, the exact replication of the facelifted terminal payoff does not always match the constrained super-replication of the payoff. The purpose of the next section is to investigate the conditions one should impose on the model dynamics $\sigma$ and the convex set $K$, in order to retrieve this useful property.

\subsection{The main result: general convex constraints} 

 We now consider general {\it Delta} constraints characterized by a subset $K$ of $\R^d$ satisfying the following assumption :

\begin{center}
 $\HYP{_K}$ $\qquad$ $K$ is a closed convex subset of $\R^d$ with non empty interior and $0 \in K$.\\
\end{center}

 We consider a stocks' price process $X$ with general dynamics \reff{dynamicsRN}. The next theorem provides a tractable necessary and sufficient condition ensuring that, in order to super-replicate under $K$-constraints any option with payoff function satisfying \HYP{h}, one simply needs to replicate the facelift transform of  the terminal payoff. 
 \vspace{2mm}
 
For any point $y$ on the boundary  $\partial K$ of $K$, we denote by $ N_K(y)$ the set of unitary outward normal vectors to $K$ at $y$ i.e. 
 \b*
 N_K(y) & := & \Big\{ n\in \R^d~:~|n|=1~\mbox{ and }~  n^\top(y-y')  \le 0 \quad \forall y'\in K \Big\}\;.
 \e* 
 We define $\breve{\partial K}$ the set of points  $y \in \partial K$ where there exists only one outward normal vector denoted $n(y)$, i.e. 
\be\label{eq def breve K}
\breve{\partial K} &:=& \big\{y\in \partial K \;, \;\;  N_K(y)=\{n(y)\} \big\} \;.
\ee
Equivalently,  $\breve{\partial K}$ is the set of the boundary points where there is a tangent hyperplane, see \cite{roc96} for details.

{In the sequel, we shall sometimes use the following technical assumption on the couple $(\sigma,K)$. 
}

\no{ \HYP{_G} For any $(t,x)\in [0,T)\times \textrm{Supp}(\sigma)$ and any $y\in{\breve{\partial K}}$, the  $\F^t$-local martingale $(M^{t,x,y}_s)_{s\in[t,T]}$ defined by
\b*
M^{t,x,y}_s & = & \Ec_s\left(\int_0^. \left(\sigma^{-1}(X_r^{t,x})\Big(\sum_{1\leq i,k\leq d} \partial_k[\sigma\sigma^\top]^{i,j}(X_r^{t,x})n^i(y)n^k(y) \Big)_{1\leq j \leq d}\right)^\top dW_r\right)\;,
\e* 
for $s\in[t,T]$, is an $\F^t$-martingale.}

We refer to Remark \ref{remark_BS} for a discussion on relevant cases when $\HYP{_G}$ is satisfied.

\vspace{2mm}

Finally, for any $y\in \breve{\partial K}$, we associate to $n(y)$ a family  $(\bar n_\ell(y), 1\le \ell\le d)$ of $d$ vectors such that $\bar n_1(y) := n(y)$ and $(\bar n_1(y),\ldots,\bar n_{d}(y))$ is an orthonormal basis of $\R^d$.  We denote by $P(y)$ the new matrix basis i.e. $P(y)e_\ell = \bar n_\ell(y) $, $1 \le \ell \le d$. Observe that $P(y)$ is an orthogonal matrix. \\
When it is clear from context, we shall  omit the '$y$' in the above notations, for the reader's convenience.

\begin{Theorem} \label{maintheorem} 
Let us consider the two conditions:
\begin{enumerate}[(i)]
\item For any  payoff function $h$ satisfying \HYP{h} and any $(t,x)\in [0,T]\times{\textrm{Supp}(\sigma)}$, the super-replicating price and strategy of $h(X_T^{t,x})$ under $K$-constraint 
 coincides with the exact replicating price and unconstrained strategy of the facelifted claim $F_{K}[h](X^{t,x}_T)$.
\item The following holds true:
\be\label{Condition}
\partial_x[ \bar n_\ell^\top(y) \sigma  \sigma^\top(.) \bar n_k(y) ] n(y) &=& 0 
\;, \qquad  2 \le  k,\ell \le d\;,\qquad
\ee

for all  $y\in\breve{\partial K}$.
\end{enumerate}
Under $\HYP{_K}$, we have that  (i) implies to (ii). Moreover, if $\HYP{\sigma}$ and $\HYP{_G}$ hold then $(ii)$ implies $(i)$.
\end{Theorem}

In order to alleviate the presentation of the paper, the proof of this theorem is postponed to Section \ref{SectionRegu}. Considering first regular payoff functions $h$ which are stable under the facelifting procedure, the unconstrained hedging strategy of $h(X_T)$ interprets as the solution of a linear BSDE (or PDE) with terminal condition $\nabla h(X_T)\in K$. We introduce in Section \ref{SectionViability} the notion of first order viability for the corresponding BSDE which ensures that  the solution of the BSDE  is valued at any time in $K$, for any bounded terminal payoff function of the form $\nabla h(X_T)$ lying in $K$. We then establish in Theorem \ref{ThmViabilityConvex} that  Condition (ii) above is necessary and sufficient for this newly introduced first order viability property. 
The extension of this property to  payoff functions satisfying $\HYP{h}$ is done via a regularization argument presented in Section  \ref{SectionRegu}.

\begin{Remark}{\rm 
The previous property extends also naturally to American options, under additional regularity assumptions. See  Remark \ref{rm stoptime2} (ii) below for a sketch of proof. 
}\end{Remark}

\begin{Remark}\label{remark_BS}{\rm 
Let us  exhibit interesting cases where $\HYP{_G}$   holds.

\no (i) The Black and Scholes model.\\ Suppose that the volatility function $\sigma$ is given by
\be\label{sigmaBS}
\sigma(x) & = & Diag(x)\Sigma\;,\quad x\in \R^d
\ee
 where $\Sigma$ is an invertible matrix of $M_d$. 
As detailed in Example 5 below, Condition \reff{Condition} imposes that $n(y)$ is a vector of the canonical basis, for any $y\in \breve{\partial K}$. Denoting then $e_{i_0}$ the outward normal vector $n(y)$ of interest,  the relation $n^i(y)n^k(y)=\mathbf{1}_{i=k=i_0}$ for any $i,k$ together with \reff{sigmaBS} imply via a direct computation that the local martingale $M^{t,x,y}$ rewrites
 \b*
 M^{t,x,y}_s & = & \Ec_s\Big(\int_0^. \left(\Sigma^{-1}\big((1+\1_{j=i_0})[\Sigma\Sigma^\top]^{i_0,j}\big)_{1\leq j \leq d}\right)^\top dW_r\Big)\\
  & = & \Ec_s\Big(\int_0^. \Big((1+\1_{j=i_0})\Sigma^{j,i_0}\Big)_{1\leq j\leq d}^\top dW_r\Big)
 \e*
 for all $s\in [t,T]$.  Therefore, Assumption $\HYP{_G}$ is satisfied.

\no (ii) The elliptic volatility model.
\\ Suppose that there exists two constants $C_1>0$ and $C_2>0$ such that 
\be\label{sigmabdd}
|\sigma(.)| \; \leq \; C_1   \quad \mbox{and}\quad \sigma^\top\sigma(.) & \geq & C_2 I_d \;.
\ee
Observe that $\textrm{Supp}(\sigma)=\R^d$ and Assumption (\textbf{H}$\sigma$) holds. 
Using \reff{sigmabdd}, we compute
\b*
\Big|\sigma^{-1}\big(\sum_{1\leq i,k\leq d} \partial_k[\sigma\sigma^\top]^{i,j}n^i(y)n^k(y) \big)_{1\leq j \leq d}\Big|^2 
& \leq & 
{1\over C_2} \Big| \big(\sum_{1\leq i,k\leq d} \partial_k[\sigma\sigma^\top]^{i,j}n^i(y)n^k(y)\big)_{1\leq j \leq d}\Big|^2\\ 
& \leq & 
{d^3\over C_2} \sup_{\R^d}|\partial_.[\sigma\sigma^\top]| \,.
\e*  
Since $\sigma$ is $C^1_b$ and bounded, we deduce that the term above is bounded, and using Novikov condition that, $\HYP{_G}$ holds.

}\end{Remark}

We conclude this section with the following Remark discussing equivalent writing of condition \eqref{Condition}.

\begin{Remark}\label{remcondbasefree}{\rm  
(i) An equivalent coordinate-free formulation of condition \reff{Condition} is the following:
\begin{align}\label{condbasefree}
\partial_x \Big[ \text{{\rm Tr}} \big( \sigma \sigma^\top(x)  \gamma \big ) \Big] n(y) = 0 \;, \quad \forall  (x,y,\gamma)\in \text{Supp}(\sigma)\times \breve{\partial K}\times S_d\;\;  \mbox{s.t. } \; \gamma\, n(y)=0\;.\quad
\end{align}
Indeed, fixing $(x,y)\in \text{Supp}(\sigma)\times \breve{\partial K}$, observe that the family $(\epsilon_{k\ell})_{1 \le k \le \ell \le d}$ of elements of $S_d$, given by
 \b*
 \epsilon_{k\ell} &:= \bar n_\ell(y) \bar n_k(y)^\top  + \bar n_k(y) \bar n_\ell(y)^\top  \;=\;  P(y)(e_\ell e_k^\top + e_k e_\ell^\top )P(y)^\top \;, \quad 1 \le k  \le \ell \le d
 \e*
  is a basis of $S_d$.  Moreover, it is straightforward to show that the family $(\epsilon_{k\ell})_{2\le k \le \ell \le d}$ is a basis of  $\{\gamma\in S_d\;, s.t. \;\;\;  \gamma n=0\}$. Thus the relation
  \b*
  \partial_x \Big[ \text{{\rm Tr}} \big( \sigma \sigma^\top(x)  \epsilon_{k\ell}  \big ) \Big] n(y) = \partial_x[ \bar n_\ell^\top \sigma  \sigma^\top \bar n_k ] n(y)  \;, \quad 1 \le k  \le \ell \le d
  \e*
   implies that \reff{Condition} and  \reff{condbasefree} are equivalent.
\\
\vspace{2mm}
\\
\no
(ii)  Condition \eqref{Condition} can also be rewritten fully in the new basis $(\bar n_\ell(y))_{1 \le \ell \le d}$, for a fixed $y\in\breve{\partial K}$. Let us define
$\tilde \sigma(\cdot ) := P^\top\sigma(P\cdot )$, then  \eqref{Condition} simply reads
\begin{align*}
\partial _1 \big [ \tilde \sigma^{k.} (\tilde \sigma^{\ell.})^\top \big ]  = 0 \;,
\end{align*}
for all $2 \le k \le \ell \le d$. If we define $\tilde X := P^\top X$, then the above condition states that there is no dependency upon the first component of $\tilde X$ of the quadratic covariations of the other components. 
}\end{Remark}

\subsection{Financial applications}

We now present financial applications of the main result of the paper. We also precise the form of the necessary and sufficient condition \reff{Condition} for relevant convex constraints and model dynamics in the field of mathematical finance. We successively consider the cases of illiquid assets, short sell prohibition and restrictions on the total number of positions taken on the financial market. 

Then, we look towards the  model dynamics satisfying the viability property for any possible closed convex constraints set. Although it is always the case in dimension $d=1$, it appears that this condition is very restrictive  in greater dimension. 
Finally, for the multidimensional Black Scholes model, we show that super-replicating an option with Delta constraint is equivalent to replicating the corresponding facelifted payoff if and only if the set of constraints is given by an hypercube $K_c$, recalling \eqref{eq de Kc}.

\no {\bf Example 1:  Non-tradable asset.}\\
In dimension 2, we consider the case where Asset $1$ is illiquid and thus cannot be traded. The corresponding convex set $K$ is the ordinate axis $\{0\}\times\R$. This convex set does not satisfy  Assumption \HYP{_K} since it has an empty interior, but Remark \ref{rm hyperplane1} below justifies that Theorem \ref{maintheorem} is also valid for hyperplanes. The only outward normal vectors to $K$ are $n=(1,0)$  and $n=(-1,0)$ which lead to the same Condition \reff{Condition}, which rewrites
\be\label{cond_illiq}
\partial_1 \left[ |\sigma^{21}|^2+|\sigma^{22}|^2\right] &=& 0 \;.
\ee
This necessary and sufficient condition indicates that the quadratic variation of the second asset does not depend on the first one. Observe that the condition derived by \cite{bqr00} for classical viability property rewrites:  $\partial_1 \sigma^{21} = \partial_1\sigma^{22} = 0$. This condition is stronger than \reff{cond_illiq} as expected.



\no {\bf Example 2:  No short sell.} \\
Consider a market with two assets where  short selling Asset $1$ is prohibited i.e. $K=\R^+\times\R$. Up to their sign, this convex set shares the exact same outward  normal vectors with the one considered in the previous example. The main observation here is that Condition \reff{Condition} is only related to the border of $K$ so that convex sets with similar borders share the same viability property. Therefore the prohibition of short selling asset $1$ is also related to Condition \reff{cond_illiq}. Moreover, since the corresponding convex sets share the same unit outward normal vectors, we emphasize that restricting to portfolios $\Delta$ such that $\Delta^1\in[-a,b]\cap\R$ for some $a, b \in [0,\infty]$ leads to the same condition \reff{cond_illiq} derived here when $(a,b)=(0,\infty)$.

Similarly, if short selling any of the two assets is prohibited, super-replication reduces to hedging the facelifted claim payoff whenever the stock model satisfies
\be\label{cond_nss}
\partial_1 \left[ |\sigma^{21}|^2+|\sigma^{22}|^2\right] \;=\; 0 \qquad &\mbox{and} & \qquad
\partial_2 \left[ |\sigma^{11}|^2+|\sigma^{12}|^2\right] \;=\; 0 \;.
\ee
In dimension $d$, when the subset $J\subset\{1,\ldots,d\}$ of assets cannot be short sold,  the necessary and sufficient Condition \reff{Condition}  rewrites as a constraint on quadratic covariations and takes the following form:
\b*
\partial_j \left[ \sigma^{\ell 1}\sigma^{k 1} + \cdots + \sigma^{\ell d}\sigma^{k d} \right] &=& 0 \;, \qquad  j\in J\;, \;\;\; \ell, k \in\{1,\ldots,d\}\setminus \{j\}\;.
\e*

\no {\bf Example 3:  Bound on the number of shares.} \\
 We now consider  the case where the investor faces a constant upper bound $C$ on the total number of possible positions he or she can take on the financial market. In dimension $2$, this corresponds to the consideration of the lozenge convex set $K=\{(\Delta^1,\Delta^2)\in\R^2\;/\; |\Delta^1|+ |\Delta^2| \le C\}$. Up to their directions, there are  two outward  normal vectors for the convex set $K$: namely $(1,1)$ and $(-1,1)$. Thus, the condition \reff{Condition} rewrites as 
\b*
\left\{
\begin{tabular}{lll}
$\partial_1 \left[ |\sigma^{11} - \sigma^{21}|^2 + |\sigma^{12} - \sigma^{22}|^2 \right]
+ 
\partial_2 \left[ |\sigma^{11} - \sigma^{21}|^2 + |\sigma^{12} - \sigma^{22}|^2 \right]$
 &$=$& $0$\\
$\partial_1 \left[ |\sigma^{11} + \sigma^{21}|^2 + |\sigma^{12} + \sigma^{22}|^2 \right]
- 
\partial_2 \left[ |\sigma^{11} + \sigma^{21}|^2 + |\sigma^{12} + \sigma^{22}|^2 \right]$
 &$=$& $0$
 \end{tabular}
 \right.\;.
\e*
Observe that this condition  is the one given by \reff{cond_nss}, but  simply written in a different  orthonormal basis.

\noindent
 {\bf Example 4:  Model with unconditional viability property.}\\
Clearly, Condition \reff{Condition} is always satisfied if the volatility function $\sigma$ is constant. Therefore, the replication property (i) of Theorem \ref{maintheorem} is valid for any closed convex set in the particular  case where the assets' price  $X$ is a multidimensional Brownian motion, i.e. in the Bachelier model.\\
Moreover, in dimension 1, Proposition \ref{Propo31} states that any model satisfies this property. \\
We now show that this unconditional viability property may lead to strong restriction on the model in dimension greater than $1$.
To this end, we identify the models with separate dynamics and invertible volatility matrix satisfying the unconditional viability property in dimension 2. 
Namely, the model has the form
 \b*
 d X^i_t &=& \sigma^{i1} (X^i_t) dW^1_t + \sigma^{i2} (X^i_t) dW^2_t \;, \qquad 0\le t \le T \;, \qquad i=1,2\;.
 \e*
 For any outward normal vectors $(a,b)\in\R^2$, Condition \reff{Condition} rewrites 
 \b*
a \partial_1 \left[ |a\sigma^{21}-b\sigma^{11}|^2 + |a\sigma^{22}-b\sigma^{12}|^2\right] + b \partial_2 \left[ |a\sigma^{21}-b\sigma^{11}|^2 + |a\sigma^{22}-b\sigma^{12}|^2\right] &=& 0\;.
 \e*
 Hence, a model $\sigma$ satisfies the replication property for any closed convex set if and only if 
 \b*
 ab \left( \partial_2\sigma^{21}-\partial_1\sigma^{11} ~,~  \partial_2\sigma^{22}-\partial_1\sigma^{12} \right)  \sigma^\top (-b, a)^\top &=& 0\;, \qquad (a,b)\in\R^2 \;.
 \e*
 Since the volatility function $\sigma$ is invertible, this condition is equivalent to the relation 
 \b*
 \partial_1 \sigma^{11} \;=\;  \partial_2 \sigma^{21} &\quad\mbox{ and }\quad&  \partial_1 \sigma^{12} \;=\;  \partial_2 \sigma^{22}\,. 
 \e*

Then, the volatility function reads
\begin{align*}
\sigma(x) = \left(
\begin{array}{cc}
\gamma x_1 + \Sigma^{11} & \nu x_1 + \Sigma^{12}
\\
\gamma x_2 + \Sigma^{21} & \nu x_2 + \Sigma^{22}
\end{array}
\right)
\end{align*}
for some $\nu, \gamma \in \R$ and $\Sigma \in M_2$.
\\
The invertibility condition on $\sigma$ imposes $\nu=\gamma=0$ and $\Sigma$ invertible. Necessarily, the class of model with separate dynamics and invertible volatility matrix satisfying an unconditional viability property  is the class of  Bachelier models.

\no
{\bf Example 5:  The multidimensional Black Scholes model.}\\
We assume that the dynamics of the stocks are given by
 \b*
 d X_t &=&  Diag(X_t) \Sigma dW_t \;, \qquad 0\le t \le T  \;,
\e* 
with $ \Sigma\in M_d$ is invertible.\\
1. If the property is satisfied for a given convex set $K$, then $K$ is an hypercube. 
\\Indeed, Remark \ref{remcondbasefree} (i) yields 
\begin{align*}
\Big[\partial_x\textrm{Tr}\big(Diag(x)\Sigma\Sigma^\top Diag(x)  \gamma \big ) \Big] n(y) = 0 \;, \quad \forall  (x,y,\gamma)\in \text{Supp}(\sigma)\times \breve{\partial K}\times S_d\;\; \mbox{s.t. } \; \gamma n(y)=0\;.
\end{align*}
Since $\gamma$ and $\Sigma\Sigma^\top$ belong to $S_d$, this rewrites
\begin{align}\label{condCgamma}
\sum_{i=1}^d [\Sigma\Sigma^\top]^{i,j}\gamma^{i,j}n^i(y) & =  0  \;, \;\; 1\le j \le d\,,\; \quad  \quad \forall  (y,\gamma)\in  \breve{\partial K}\times S_d\;\; \mbox{s.t. } \; \gamma n(y)=0\;.
\end{align}
If we define $\gamma_0\in S_d$ by 
\begin{equation*}
\gamma_0  ~ = ~ \left(
\begin{array}{ccccc}
-|n^2(y)|^2 & n(y)^1n(y)^2 & 0 & \cdots & 0\\
 n(y)^1n(y)^2  & |n^2(y)|^2 & 0 & &  \vdots\\
 0  & 0 & 0 & &\vdots \\
  \vdots  & & & \ddots & \vdots \\
  0    & \cdots &  \cdots &  \cdots &0 
\end{array}
\right)
\end{equation*} 
we easily check that  $\gamma_0 n=0$. Then using \reff{condCgamma}, we get  
\b*
 [\Sigma\Sigma^\top]^{1,1}|n^{1}(y)n^2(y)|^2  & = &  [\Sigma\Sigma^\top]^{1,2}|n^{1}(y)n^2(y)|^2 \\
{[\Sigma\Sigma^\top]}^{2,1}|n^{1}(y)n^2(y)|^2 & = & [\Sigma\Sigma^\top]^{2,2}|n^{1}(y)n^2(y)|^2 
\e*
 which gives $n^{1}(y)n^2(y)=0$ since $\Sigma$ is invertible. Using a similar argument, we prove that  $n^{i}(y)n^j(y)=0$ for $i\neq j$, which shows that $n(y)$ must be a vector of the  canonical basis $(e_1,\ldots,e_d)$.\\
2. If $K=K_c$, we can apply Proposition \ref{Prop_2} to obtain the first-order viability, when the terminal condition is smooth enough. For the general case, we apply our main result. Indeed, Remark \ref{remark_BS} implies that $\HYP{_G}$ holds for the couple $(\sigma,K)$. The assumptions of Theorem \ref{maintheorem} (ii) are thus satisfied.  Setting, w.l.o.g $n(y)=e_1$, we get that Condition \eqref{Condition} holds since it rewrites in this case
 \b*
 \partial_x \Big[ \text{{\rm Tr}} \big( \sigma \sigma^\top(\cdot ) e_\ell (y) e^\top _k (y) \big ) \Big] n(y) &=& \partial_1 \left(\sum_{j=1}^d \sigma_{lj} \sigma_{kj} x_l x_k \right) \;=\; 0 
 \;, \qquad  2 \le  k,\ell \le d\;.\qquad
\e*


%
%

\section{First order viability for the PDE $\Lc_\sigma u=0$} \label{SectionViability}



In this section, we introduce a new notion of viability for linear PDEs. 
Namely, the constrained super-replication problems considered in this paper entails to find out whether a hedging strategy with terminal value in $K$ will always do so on the time interval $[0,T]$. Since the hedging strategy rewrites as the gradient of a function of $X$ solving the linear PDE $\Lc_\sigma u=0$, we want to know if the gradient of the solution of this PDE lies in $K$ on $[0,T]$ as soon as it does so at time $T$. 
This leads to the notion of first order viability property for the PDE $\Lc_\sigma u=0$,  presented in Section \ref{subsec viab_def}.  This notion interprets also in terms of viability for  linear BSDEs associated to the subclass of gradient type terminal functions. We verify in Section \ref{Subsec viab_halfspace} that the PDE $\Lc_\sigma u=0$ is first order viable for a  closed convex set $K$ with non empty interior whenever it is first order viable for almost all its  tangent half-spaces. Specializing then our study on first order viability for half-spaces, we provide 
in Section \ref{Subsec viab_hs} a geometric condition indicating whether or not the PDE $\Lc_\sigma u=0$ is first order viable for a given half-space. This finally allows us to verify in Section \ref{Subsec viab_main} that the PDE $\Lc_\sigma u=0$ is first order viable for a closed convex set $K$ if and only if the structural condition \reff{Condition} on the couple $(\sigma,K)$ is satisfied.

\subsection{First order viability: PDE and BSDE viewpoints}\label{subsec viab_def}

For any $h \in C^1_b(\R^d,\R)$,  the price function $u^h$ of the European option with terminal payoff function $h$ and maturity $T$ is given by:
 \b*
 u^h(t,x) &:=&\esp{h(X^{t,x}_T)} \,, \qquad (t,x)\in[0,T]\times\R^d\;.
 \e*
 This function interprets as  the unique viscosity solution of the parabolic PDE
\begin{equation}\label{EDPparab-uh}
\left\{\begin{array}{rccl}
\Lc_\sigma u(t,x) & = & 0  & \mbox{ for } (t,x)\in[0,T)\times\R^d\;,\\
u(T,x) & = & h(x) & \mbox{ for } x\in\R^d\;,
\end{array}\right.
\end{equation}
in the class of continuous functions with polynomial growth, see e.g. \cite{pham09}.
Moreover, by Theorem 3.1\footnote{An uniform ellipticity condition for $\sigma$ appears in the statement of this theorem but this assumption is not used in the proof and indeed not required.} in \cite{majzha02}, we deduce that $u^h$ is $C^{1}_b$ on $[0,T]\times\R^d$.

We now introduce the notion of first order viability. 
 
 \begin{Definition} \label{de 1st order viab}
 The PDE $\Lc_\sigma u=0$ is first order viable for a given set  $C$ (or $C$-first order viable) if and only if, for any function $h\in C^1_b(\R^d,\R)$ s.t.
 \b*
 \nabla h (x) & \in & C \;, \qquad x\in\R^d\;,
 \e*
 the function $u^h$ defined by \reff{EDPparab-uh} satisfies 
  \b*
 \nabla u^h (t,x) & \in & C \;, \qquad  (t,x)\in[0,T]\times\R^d\;.
 \e*
 \end{Definition}
 
 \begin{Remark}
 {\rm Contrary to the classical definition of viability, which requires the function $u^h$ to take  value in $C$, our definition deals with the \emph{first derivative}  of $u^h$ which has to be valued in $C$. 
 }
 \end{Remark}

The first order viability property for the PDE also has a direct interpretation in terms of linear BSDE solution. Indeed, as detailed in the next proposition, $\partial_x u^h (t,x)$ admits a BSDE representation, for any $(t,x)\in[0,T]\times\R^d$ and $h\in C^1_b(\R^d,\R)$.

 For $t\in[0,T]$, we denote  by $\Sc^2[t,T]$  the set of $\F^t$-adapted continuous processes $(U_s)_{s\in[t,T]}$ valued in $\R^d$ and by  $\Hc^2[t,T]$ the set of  $\F^t$-predictable processes $(V_s)_{s\in[t,T]}$ valued in $M_{d}$ such that
\b*
\E\Big[\sup_{s\in[t,T]}|U_s|^2\Big] \; < \; \infty\; \quad \mbox{and} \quad \;  \E\Big[\int_{t}^T|V_s|^2ds\Big] \; < \; \infty\;.
\e*

\begin{Proposition} \label{pr representation} Let $h$ be in $C^1_b(\R^d,\R)$. For $(t,x)\in[0,T]\times \R^d$, we consider the process
 $(\Delta^{t,x,h},\Lambda^{t,x,h})\in\Sc^2[t,T]\times\Hc^2[t,T]$  solution of the BSDE
\be\label{BSDE}
\Delta^{t,x,h}_s & = & \nabla h(X^{t,x}_T)+\int_s^TF_\sigma(X^{t,x}_r,  \Lambda_r^{t,x,h})dr-\int_s^T\Lambda_r^{t,x,h}dW_r\;,\quad s\in[t,T]\;,
\ee
where $F_\sigma:~\R^d\times M_d\rightarrow\R^d$ is defined by
\be\label{defgenf}
F_\sigma(x,\Lambda) & : = & 
\sum_{j=1}^d[\partial_x\sigma^{.j}(x)]^\top\Lambda^{.j} \;, \quad   x\in\R^d\;, \quad \Lambda\in M_d\;.
\ee
Then we have 
\b*
\Delta_s^{t,x,h} & = & \nabla u^h(s,X^{t,x}_s)\;,\qquad s\in[t,T]\;, \quad x\in\R^d\;.
\e*
Moreover, under \HYP{\sigma}, we have that $ \Lambda^{t,x,h} = \Gamma^{t,x,h} \sigma(X^{t,x})$, for some symetric matrix valued process $\Gamma^{t,x,h}$.
\end{Proposition}

 Observe that the first order viability for the PDE $\Lc_\sigma u=0$ can be directly rewritten in terms of (zero-order) viability property for the linear BSDE \reff{BSDE} on a subclass of gradient payoff functions. 

\begin{Corollary} \label{cor viab BSDE} The PDE $\Lc_\sigma u=0$ is first order viable for a given set $C$ if and only if, for any $h\in C^1_b(\R^d,\R)$ such that $\nabla h$ belongs to $C$, the part $(\Delta^{t,x,h}_s)_{t\le s \le T}$ of the solution of the BSDE \reff{BSDE} belongs $\P-$a.s. to $C$, for any $(t,x)\in[0,T]\times\R^d$. 
\end{Corollary}

\noindent \textbf{Proof of Proposition \ref{pr representation}.} Fix $(t,x)\in[0,T]\times\R^d$ and $h\in C^1_b(\R^d,\R)$. 
 Let $(Y^{t,x},Z^{t,x})\in\Sc^2[t,T]\times\Hc^2[t,T]$ be the solution to the following  BSDE with no driver:
\b*
Y^{t,x}_s & = & h(X^{t,x}_T)-\int_s^T Z^{t,x}_rdW_r\;,\quad s\in[t,T]\;.
\e*
Such a solution exists and is unique since $h\in C_b^1(\R^d,\R)$ and $\sigma\in  C^1_b(\R^d, M_d)$. 
We consider the inverse of the tangent process $\partial_xX^{t,x}$  as well as the tangent process $(\partial_xY^{t,x},\partial_x Z^{t,x})$ of $(Y^{t,x},Z^{t,x})$, see e.g. \cite{majzha02}. They have the following dynamics
\begin{align*}
[\partial_x X^{t,x}_s]^{-1}  &=  I_d+\int_t^s[\partial_x X^{t,x}_r]^{-1}\big( 
\sum_{j=1}^d[\partial_x\sigma^{.j}(X_r^{t,x})]^2 \big)dr
  -\sum_{j=1}^d\int_t^s[\partial_x X^{t,x}_r]^{-1}[\partial_x\sigma^{.j}(X_r^{t,x})]dW^j_r\;,
\\
\partial_xY^{t,x}_s & =  \partial_xh(X^{t,x}_T)\partial_x X^{t,x}_T-\sum_{j=1}^d\int_s^T\partial_xZ_r^{t,x,j}dW^j_r\;,\qquad t\le s\le T\;.
\end{align*}
From e.g. Theorem 3.1 in \cite{majzha02}, we know that 
\begin{align}
\partial_x u(t,x) = \partial_xY^{t,x}_x \; . \label{eq temp pr identification}
\end{align}

Recalling that the process $(\Delta_s^{t,x})_{t\le s \le T}$ is solution to the markovian linear BSDE \eqref{BSDE} with continuous coefficient function, we have that $\Delta_s^{t,x} = v(s,X^{t,x}_s)$ for some continuous function $v$, see e.g. Theorem 4.1 in \cite{karpen97}. Observe also that $\Delta_t^{t,x} = v(t,x)$ is deterministic.
Applying It\^o's formula, we compute using the dynamics of  $[\partial_x X^{t,x}_s]^{-1}$ and $\partial_xY^{t,x}$  that 
\begin{align}\label{eqtemp1212}
[\Delta_s^{t,x}]^\top  \;=\; \partial_x Y_s^{t,x}[\partial_xX^{t,x}_s]^{-1}  \;,\qquad t\le s \le T\;.
\end{align}
Setting $s=t$ in the above equation and using \eqref{eq temp pr identification}, we obtain that $\Delta^{t,x}_t = [\partial_x u]^\top(t,x)$. 
\\
We deduce that $\Delta^{t,x}_s = \nabla u(s,X^{t,x}_s)$, for any $s\in[t,T]$.

When the volatility matrix $\sigma$ is smooth, one can show, using Feynman-Kac formula, that $\Delta^{t,x,h}_t$ is a classical solution of a linear PDE and then $\Lambda^{t,x,h}_t = \partial_{xx}u(t,x) \sigma(x)$. In the general case, one uses a regularization procedure (see e.g. the proof of Proposition 3.3 in \cite{boucha08}) and the stability property of (linear) BSDEs (see e.g. Proposition 2.1 in \cite{karpen97}) to show that $\Gamma^{t,x,h}$ is the limit of symetric matrix and thus symetric itself.
\ep

The rest of the section is dedicated to the proof of a necessary and sufficient condition for the first order viability property to hold. It is important to observe that the only possible terminal conditions for the linear BSDE \reff{BSDE} are of the form $\nabla h(X_T)$. Hence the viability characterization for BSDEs derived in \cite{bqr00} does not apply directly here since it requires the consideration of any terminal condition of the form $g(X_T)$, with $g$ a continuous function. In the next section, we adapt the arguments developed in  \cite{bqr00} using a geometric approach.

\subsection{First order viability and half-space decomposition}\label{Subsec viab_halfspace}
In this section, we prove that the first order viability for a closed convex set $K$ satisfying $\HYP{_K}$, is characterized by the first order viability of a well chosen collection of  half-spaces $H$ supporting $K$, i.e. such that $K\subset H$ and $\partial H\cap K\neq \emptyset$. For this purpose, we first rewrite $K$ as the intersection of the corresponding half-spaces and then discuss the related first order viability properties. 

 For $y\in \partial K$, we denote by $\rho(y)$ the radius of the largest closed ball included in $K$ which is tangent to $K$ at point $y$ i.e. 
\b*
\rho(y) &:=& \sup\left\{ \rho \ge 0 \;|\; y\in \bar{B}(x,\rho) \subset K\;, \quad \mbox{for some $x\in K$}  \right\} \;.
\e*
 The set of points of  $\partial K$ with $\rho>0$ corresponds to a subset of points where the convex surface presents some regularity. We denote this subset by $\tilde{\partial K}$, namely,
  \b*
 \tilde{\partial K} &:=& \left\{ y\in\partial K\;, \quad \rho(y)>0 \right\} \;.
 \e*
 In particular, observe that for any point in $\tilde{\partial K}$, there exists a unique outward  normal vector so that
  $ \tilde{\partial K} \;\subset\; \breve{\partial K} \;\subset\; \partial K,$
 where $\breve{\partial K}$ is defined in \reff{eq def breve K}. 
 For $y\in  \tilde{\partial K}$, we denote by $n(y)$ the unique outward  normal vector and by $H_y$ the half-space tangent to $K$ at point $y$ containing $K$, i.e.
 \b*
 H_y &:=& \left\{ y'\in\R^d\,, \;\; (y'-y)^\top {n}(y) \le 0 \right\}  \;, \qquad y\in \tilde{\partial K}\;.
 \e*
 
 \begin{Lemma}\label{LemmaDecompConv}
 Any  convex set $K$ satisfying \HYP{_K} rewrites
 \be\label{convexdecomposition}
  K  &= &\; \bigcap_{y\in \tilde{\partial K}} H_y \;.
 \ee
 \end{Lemma} 
 
 \proof The proof divides in two steps.

\noindent \textbf{\textit{Step 1:}} Theorem 18.8 in \cite{roc96} states that $x \in K$ if and only if
 \begin{align}
(x-y)^\top n(y)\; \le \; 0\,,\; \quad y \in  \breve{\partial K} \;. \label{eq tangent}
 \end{align}
 Let $A$ be a dense subset of $\breve{\partial K}$. It is obvious that $x \in K$ implies
 \begin{align}
(x-y)^\top n(y) \; \le \; 0\,,\quad  y \in  A \;. \label{eq tangent special}
 \end{align}
 We are going to verify the converse implication, showing that \eqref{eq tangent special} implies \eqref{eq tangent}.
 For this to be true, we only need to find for any fixed $y \in \breve{\partial K}$ an approximating sequence $(y_i)$ of points in $A$ s.t. $y_i \rightarrow y$ and $n(y_i) \rightarrow n(y)$. 
\\
Let first observe that for $y \in \breve{\partial K}$, since $A$ is a dense subset of $\breve{\partial K}$, there exists  
$(y_i)_i$  an approximating sequence of points in $A$ converging to $y$.  
Since $\set{n(y_i), i \ge 0}$ is compact, we have that, up to a subsequence still denoted $(y_i)$, $(n(y_i))_i$ converges to some unit vector $\nu$. Moreover,  we compute 
 \begin{align*}
 0 \ge  (x-y_i)^\top n(y_i) =  (x-y)^\top n(y_i)  +  (y-y_i)^\top n(y_i)  \;, \qquad x\in K\;, \qquad i\ge 1\;.
 \end{align*}  
 This implies that $ (x-y)^\top \nu \le 0$ for any $x\in K$, so that $\nu$ is an outward  normal vector for $K$ at $y$. Since $y\in \tilde{\partial K}$, we get $\nu = n(y)$ and \reff{eq tangent special} holds.
 
\noindent \textbf{\textit{Step 2:}}  Let $B$ be the unit closed ball and set $K^\alpha = K \cap \alpha B$ for $\alpha\in\N$, with $\alpha\ge 1$. Hence $K=\cup_{\alpha} K^\alpha$. For $\alpha\ge 1$, one observes that if $x\in \breve{\partial K} \setminus \tilde{\partial K}$ and $|x|<\alpha$, then $x\in\breve{\partial K^\alpha} \setminus \tilde{\partial K^\alpha}$. McMullen \cite{mcm74} shows that $\breve{\partial K^\alpha} \setminus \tilde{\partial K^\alpha}$ has  a zero $\R^{d-1}$-lebesgue measure, for any $\alpha\ge1$.  Since $K=\cup_{\alpha} K^\alpha$, this implies that  $\tilde{\partial K}$ is dense in $\breve{\partial K}$. Combined with Step 1, this concludes the proof of the lemma.
\eproof
  
\vspace{2mm}  
  We now observe that the first order viability property for the convex set $K$ relates to the first order viability property for every half-space $H_y$, $y \in \tilde{\partial K}$. This nice property allows us to restrict  our upcoming argumentation to the consideration of viability property for half-spaces. 
  \begin{Proposition}\label{prop-viab-hyperplan}
 Let \HYP{_K} be in force. 
 Then, the PDE $\Lc_\sigma u$ is first order viable for the closed  convex set $K$ if and only if it is first order viable for every half-space $H_y$, for $y\in \tilde{\partial K}$. 
 \end{Proposition} 

The proof of this proposition requires the following lemma, which states the homothetic stability of the first order viability property.
 \begin{Lemma}\label{lem-viab-homothetie}
The PDE $\Lc_\sigma u$ is first order viable for a closed set $C$ if and only if it is first order viable for every closed set  $\lambda C + y$ with $\lambda>0$ and $y\in\R^d$.
\end{Lemma}
\proof We fix a set $C$ and choose $y\in\R^d$ and $\lambda>0$. We suppose that the PDE $\Lc_\sigma u$ is first order viable for $C$ and simply need to verify that it is also first order viable for $\lambda C + y$. Let $h \in C^1_b(\R^d,\R)$ such that 
$\partial_x h^\top$ is valued in $\lambda C + y$ and define the function $g\in C^1_b(\R^d,\R)$ by 
\b*
g(x)  & := & {1\over \lambda}\Big(h(x)- x^\top y\Big)\;, \quad  x\in\R^d\;.
\e*
The gradient of $g$ directly satisfies
\b*
\partial_x g (x)^\top  & = & {1\over \lambda}\Big(\partial_x h (x)^\top - y \Big)~\in~C\;, \quad  x\in\R^d\;.
\e*
Since the PDE $\Lc_\sigma u$ is viable for $C$ we deduce that $[\partial_x u^g]^\top$ is valued in $C$. 
Moreover, we easily check that the function $(t,x)\mapsto\lambda u^g(t,x)+  x^\top  y$ solves the PDE
\be\label{EDPparab-ug-modifie}
\Lc_\sigma u \; = \; 0   \mbox{ on } [0,T)\times\R^d\;, \quad & \quad & \quad
u(T,.) \; = \; h  \mbox{ on } \R^d\;.
\ee
From uniqueness of the solution to  \reff{EDPparab-ug-modifie}, we get 
\b*
u^h(t,x) & = & \lambda u^g(t,x)+ x^\top y\;,\quad (t,x)\in[0,T]\times\R^d\;.
\e*
Since $[\partial_x u^g]^\top$ is valued in $C$, we deduce that $[\partial_x u^h]^\top$ is valued in $\lambda C + y$. The arbitrariness of $h$ indicates that  $\Lc_{\sigma} u$ is first order viable for $\lambda C + y$ and concludes the proof.
\ep

We now to turn to the proof of Proposition \ref{prop-viab-hyperplan}.

\noindent \textbf{Proof of Proposition \ref{prop-viab-hyperplan}.} 
The proof divides in two steps, corresponding to each implication.

\textbf{\textit{Step 1:}}  Assume that the PDE $\Lc_\sigma u=0$ is first order viable for any $H_y$, with $y\in \tilde{\partial K}$. We deduce from this viability property and the representation of $K$ given in Lemma \ref{LemmaDecompConv} that, for any $h \in C^1_b(\R^d,\R)$ with $\partial_x h^\top$ valued $K$, $[\partial_x u^h]^\top$ is valued in every $H_y$, $y\in \tilde{\partial K}$. Using again the representation of $K$ given in Lemma \ref{LemmaDecompConv}, we conclude that $K$ is first order viable.

 \vspace{1mm}
 \noindent \textbf{\textit{Step 2:}}  Assume that the PDE $\Lc_\sigma u=0$ is first order viable for $K$. 
 
 We intend to prove that it is also first order viable for any $H_y$ with $y\in \tilde{\partial K}$. Up to considering $K-\{y\}$ according to Lemma \ref{lem-viab-homothetie}, we suppose that $y=0$. 
\\ 
By definition of $\tilde{\partial K}$ and denoting by $n$ the outward normal vector to $K$ at point $0$, there exists $R_0>0$ such that the ball $\bar{B}(-R_0 n,R_0)\subset K$ is tangent to $K$ at $0$. We pick $h \in C^1_b(\R^d,\R)$ with $\partial_x h^\top$ valued in $H_0$. We choose any arbitrary $\eps>0$ and intend to prove that $[\partial_x u^h]^\top$ is valued in $H_0+\eps n$.

\begin{figure}[!htb]
\begin{center}
   \centerline{\resizebox{6cm}{!}{{\includegraphics[width=\textwidth]{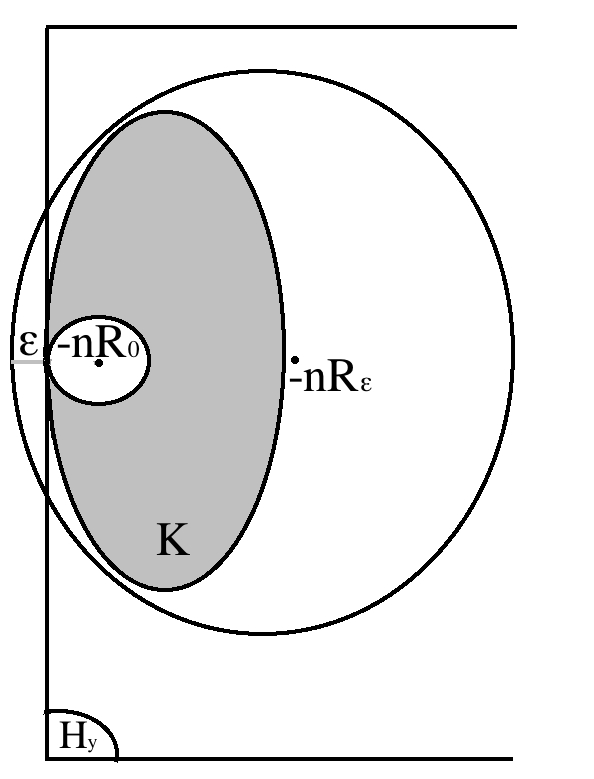}}}}
\caption{Visual representation of Step 2 in the proof of Proposition \ref{prop-viab-hyperplan}}
\label{Figure 2}
\end{center}
\end{figure}

 Since $\partial_x h^\top$ is bounded, there exists $R_\eps$ such that $\partial_x h^\top$ is valued in $\bar{B}((\eps-R_\eps)n,R_\eps)$, as shown on Figure \ref{Figure 2}. Therefore, we deduce that
 \b*
 \partial_x h^\top(x)\in \bar{B}((\eps-R_\eps)n,R_\eps) = \frac{R_\eps}{R_0} \bar{B}(-R_0 n,R_0) + \eps n \subset \frac{R_\eps}{R_0} K  + \eps n\;, \quad x\in\R^d\;.
  \e*
 Since the PDE $\Lc_\sigma u$ is first order viable for $K$, Lemma \ref{lem-viab-homothetie} indicates that it is also first order viable for $\frac{R_\eps}{R_0} K  + \eps n$ and therefore $[\partial_x u^h]^\top$ is valued in $\frac{R_\eps}{R_0} K  + \eps n$. But, since the half-space $H_0$ is convex cone, we have
 \b*
  \frac{R_\eps}{R_0} K  + \eps n
  &\subset &
  \frac{R_\eps}{R_0} H_0  + \eps n
  \;=\;
H_0  + \eps n \;.
  \e*
  Thus $[\partial_x u^h]^\top$ is valued in $H_0  + \eps n$, for any $\eps>0$. Hence, $[\partial_x u^h]^\top$ is also  valued in $H_0$. Therefore, the PDE $\Lc_\sigma u=0$ is first order viable for any hyperplane $H_y$, with $y\in \tilde{\partial K}$. \ep

\subsection{First order viability property for half-spaces}\label{Subsec viab_hs}

The aim of this section is to prove the following result.
 \begin{Theorem}\label{ThmViabilityHalfspace}
If the PDE $\Lc_\sigma u=0$ is first order viable for the half-space $H$ with normal unit vector $n$ then \be\label{EqViabilityHalfspace}
 n^\top F_{\sigma}(x,\gamma\sigma(x))  &=& 0\;, \qquad  \forall x\in{\text{Supp}(\sigma)}\;,\;\; \gamma\in S_d  \quad \mbox{ such that } \;\;\; \gamma n =0\;.
 \ee
Moreover, whenever \HYP{\sigma} and \HYP{_G} hold, the converse is valid. 
\end{Theorem}

The proof of this theorem is  done in two steps below, proving each assertion separately, namely Proposition \ref{PropCondViabNec} 
and Proposition \ref{PropCondViabSuf}. The proofs follow ideas developed in \cite{bqr00} for (zero-order) viability properties on BSDEs, but are much simpler due to the consideration of half-spaces.

 \begin{Remark}\label{rm interp}{\rm
(i) The condition $n^\top F_{\sigma}(x,\gamma\sigma(x))  \le 0$ has a natural geometric interpretation. For a given terminal condition in $H$, it indicates that the $\Delta$-component of the solution of the BSDE \reff{BSDE} with generator $F_{\sigma}$ is pushed inside the half-space $H$ whenever it comes near its boundary. This condition takes the form $n^\top F_{\sigma}(x,\gamma\sigma(x))  = 0$ in our context since $F_{\sigma}$ given in \reff{defgenf} is linear with respect to its second variable.
\\
\vspace{1mm}
\noindent
(ii) Assume $\HYP{\sigma}$ holds true. The BSDE \reff{BSDE} satisfies the classical (zero-order) viability property for the half-space $H$ if and only if 
  \b*
 n^\top F_{\sigma}(x,\gamma\sigma(x))  &=& 0\;, \qquad  \forall x\in{\text{Supp}(\sigma)}\;,\;\; \gamma\in M_d  \quad \mbox{ such that } \;\;\; \gamma^\top n  =0\;.
 \e*
This condition is the one given in \cite{bqr00} adapted to our context, it is stronger than \reff{EqViabilityHalfspace}. It is important to observe that the restriction to terminal conditions of the form $\nabla h(X^{t,x}_T)$ translates into the only consideration of symetric matrices $\gamma$.  
 }\end{Remark}

Before proving Theorem \ref{ThmViabilityHalfspace}, we state that 
Condition \eqref{EqViabilityHalfspace} and Condition \eqref{Condition} for half-spaces with outward normal vector $n$, are the same.

\begin{Lemma} \label{le cond hs}
{Condition \eqref{EqViabilityHalfspace}  is equivalent to
\begin{align}\label{eq tru cond hs}
\partial_x[ \bar n_\ell^\top \sigma  \sigma^\top(.) \bar n_k ] n &= 0 
\;, \qquad  2 \le  k \le \ell \le d\;.
\end{align}
recalling that $(n,\bar n_2,\dots,\bar n_d)$ is an orthonormal basis of $\R^d$.}
\end{Lemma}

\proof
 Recall that $P(y)$ denotes the new basis orthogonal matrix associated to $(n(y),\bar n_2(y),\ldots,\bar n_d(y))$.  We introduce the family $(\epsilon_{k\ell})_{1 \le k \le \ell \le d}$ of elements of $S_d$, given by
 \b*
 \epsilon_{k\ell} &:=& \bar n_\ell(y) \bar n_k(y)^\top  + \bar n_k(y) \bar n_\ell(y)^\top \;=\; P(e_\ell e_k^\top + e_k e_\ell^\top )P^\top\, \quad 1 \le k  \le \ell \le d\;.
 \e*
 This family is a basis of $S_d$. 
 Moreover, it is straightforward to show that the family $(\epsilon_{k\ell})_{2\le k \le \ell \le d}$ is a
 basis of  $\{\gamma\in S_d\;, s.t. \;\;\; \gamma n=0\}$.
 If we assume that 
 \begin{align} \label{eq reec F}
 F_\sigma(\cdot,\epsilon_{k\ell} \sigma) =  \partial_x[ \bar n_\ell^\top \sigma  \sigma^\top \bar n_k ]^\top\;,
 \end{align}
 it is then clear that Condition \eqref{EqViabilityHalfspace} is equivalent to Condition \eqref{eq tru cond hs}.
 \\
The following computations prove \eqref{eq reec F}:
 \b*
F_\sigma(\cdot,\epsilon_{k\ell} \sigma)
&=&
 \sum_{j=1}^d \partial_x[ \sigma^{. j}]^\top \epsilon_{k\ell} \sigma^{.j}  
 \;=\;
 \sum_{j=1}^d \partial_x[ n_\ell(y)^\top \sigma^{. j} \bar n_k(y)^\top \sigma^{.j} ]^\top \\
   &=&
\sum_{j=1}^d \partial_x[ (P^\top \sigma)^{\ell j} (P^\top \sigma)^{kj} ]^\top  
  \;=\; 
  \partial_x[ (P^\top \sigma)^{\ell .} ((P^\top \sigma)^{k.} )^\top ]^\top  \\
     &=&
 \partial_x[ e_\ell^\top P^\top \sigma (e_k^\top P^\top \sigma)^\top ]^\top 
 \;=\;  
 \partial_x[ \bar n_\ell^\top \sigma  \sigma^\top \bar n_k ]^\top
\e*
\eproof

\noindent We now proceed with the proof of the necessary part of Theorem \ref{ThmViabilityHalfspace}.
\begin{Proposition}\label{PropCondViabNec} 
If the PDE $\Lc_\sigma u=0$ is first order viable for a half-space $H$ with unit outward normal vector $n$, then Condition \reff{EqViabilityHalfspace} holds for  $n$.
\end{Proposition}

In order to derive this proposition, we use the following technical lemma whose proof is given in the Appendix, see Section \ref{Appendix3}.

\begin{Lemma}\label{lemma approx Delta}
Let  $(t,x,n,\gamma)\in[0,T]\times\R^d\times \partial \bar B(0,1) \times S_d$ such that $\gamma\,n = 0$. If the PDE $\Lc_\sigma u=0$ is first order viable for a half-space $H$ with unit outward  normal vector $n$, then  $\Delta^{t,x}$ belongs to $H$ where $(\Delta^{t,x},\Lambda^{t,x})$ is the solution on $[t,T]$ of the BSDE \reff{BSDE} associated to the terminal condition $\gamma(X^{t,x}_T-x)$. 
\end{Lemma}

\noindent\textbf{Proof of Proposition \ref{PropCondViabNec}.} 

Using Lemma \ref{lem-viab-homothetie}, we can assume w.l.o.g. that $0 \in \partial H$.
\\
In order to verify that Condition \reff{EqViabilityHalfspace} holds for the vector $n\in\partial \bar B(0,1)$, we pick any $x\in\R^d$ and $\gamma\in S_d$ such that $\gamma n  =0$. Since one can choose either $\gamma$ or $-\gamma$ and the map $\gamma \mapsto F_\sigma(x,\gamma\sigma(x))$ is linear, we only need to check that  
 \be\label{temptemp}
  n^\top F_\sigma (x,\gamma\sigma(x))
 &\le &
  0 \;.
 \ee
We pick $\eps\in[0,T]$ and denote by $X^{\eps}$ solution on $[T-\eps,T]$ of the SDE \reff{dynamicsRN} starting in $x$ at time $T-\eps$ and by $(\Delta^\eps,\Lambda^\eps)$ the solution on $[T-\eps,T]$ of the BSDE \reff{BSDE} associated to the terminal condition $\gamma(X^\eps_T-x)$. We deduce from $\gamma n =0$ that $n^top \gamma(X^\eps_T-x)=0$ so that $\Delta^\eps_T=\gamma(X^\eps_T-x)\in\partial H$. This implies that
 \b*
  n^\top \Delta^\eps_t & = &
 \int_t^T n^\top  F_\sigma (X^{\eps}_s,\Lambda^\eps_s)  ds 
 - \int_t^T n^\top  \Lambda^\eps_s   dW_s \;, \qquad T-\eps\le t \le T\;.
 \e*
Since the PDE $\Lc_\sigma u=0$ is first order viable for the half-space $H$,  Lemma \ref{lemma approx Delta} 
 indicates that $\Delta^\eps$ belongs to $H$ and thus
 \be\label{eqtempssgnndelta}
n^\top \Delta^\eps_t & \le & 0
  \;, \qquad T-\eps\le t \le T\;.
 \ee
 Let us define the process $\hat \Delta^\eps$  on $[T-\eps,T]$ by
 \b*
 \hat \Delta^\eps_t  &=& \gamma(X^\eps_t-x) +    F_\sigma (x,\gamma\sigma(x)) (T-t) \\ 
 & = & \gamma(X^\eps_T-x) +   F_\sigma (x,\gamma\sigma(x)) (T-t)  - \int_t^T \gamma\sigma(X^{\eps}_s) dW_s 
 \;, \qquad T-\eps\le t \le T\;.
 \e*
 Since $ n^\top \gamma =0$, we compute directly
 \be\label{temp12432}
 {n^\top \hat\Delta^\eps_{T-\eps}} & = & n^\top F_\sigma (x,\gamma\sigma(x)) \eps\;.
 \ee
 In particular, observe that \reff{temptemp} is satisfied as soon as ${n^\top \hat\Delta^\eps_{T-\eps}}/\eps$ is non-positive for $\eps$ small enough. \\
 Besides, $(\Delta^\eps,\Lambda^\eps)$ and $(\hat\Delta^\eps,\gamma\sigma(X^\eps))$ are solutions on $[T-\eps,T]$ of BSDEs with the same terminal condition $\gamma(X^\eps_T-x)$ and respective drivers $(F_\sigma (X_s,.))_s$ and $F_\sigma (x,\gamma\sigma(x))$. The stability property for BSDE, see for e.g. Prop 4.1 in \cite{bqr00}, reads
 \b*
 \E | \hat \Delta^\eps_{T-\eps} - \Delta^\eps_{T-\eps} |^2 
 & \le &  
 C \E\left[\left( \int_{T-\eps}^T |F_\sigma(X^\eps_s,\gamma\sigma(X^\eps_s)) - F_\sigma(x,\gamma\sigma(x))| ds \right)^2 \right] \;,
 \e* 
 where $C$  is a non-negative constant, which may change from line to line and does not depend on $\eps$. \\
The Cauchy Schwartz inequality together with the Lipschitz property of the driver with respect to its second variable leads to
 \be\label{eqtemp7512}
 \E | \hat \Delta^\eps_{T-\eps} - \Delta^\eps_{T-\eps} |^2 
 & \le &  
 C \eps \int_{T-\eps}^T  \E \left[|\gamma\sigma(X^\eps_s) - \gamma\sigma(x) |^2 +   |F_\sigma(X^\eps_s,\gamma\sigma(x)) - F_\sigma(x,\gamma\sigma(x))|^2\right] ds \nonumber\\
 & \le &  
 C \eps^2 (\eps +  \alpha_\eps)\;,
 \ee
 where the last inequality follows from classical estimates on the forward diffusion $X^\eps$ on $[T-\eps,T]$,  and $\alpha_\eps$ is given by
 \b*
 \alpha_\eps 
 &:=& 
 \E \left[\sup_{T-\eps \le s \le T }  |F_\sigma(X^\eps_s,\gamma\sigma(x)) - F_\sigma(x,\gamma\sigma(x))|^2\right] \;.
 \e*
 Observe from the Markov property of the process $X^\eps$ that $\alpha_\eps$ rewrites 
 \b*
 \alpha_\eps 
 &=& 
 \E \left[\sup_{0 \le s \le \eps }  |F_\sigma(X^{0,x}_s,\gamma\sigma(x)) - F_\sigma(x,\gamma\sigma(x))|^2 \right] \;.
 \e*
 Since $\sigma\in C^1_b(\R^d,\R^d)$, the function $F_\sigma(.,\gamma\sigma(x))$  given in \reff{defgenf} is continuous and bounded. Therefore, the continuity of the process $X^{0,x}$ together with the dominated convergence theorem ensures that $ \alpha_\eps$ goes to $0$ as $\eps$ does so. Thus, we deduce from \reff{eqtemp7512} that
  \b*
 \left\| \frac{\hat \Delta^\eps_{T-\eps}}{\eps} - \frac{\Delta^\eps_{T-\eps}}{\eps} \right\|_{L^2} 
 & \longrightarrow&  0 \; \quad \mbox{ as } \quad {\eps\rightarrow 0} \;.
 \e* 
 Up to a subsequence, this implies that $\hat \Delta^\eps_{T-\eps}/\eps$ and $\hat \Delta^\eps_{T-\eps}/\eps$ share a.s. the same limit. Therefore \reff{eqtempssgnndelta} together with \reff{temp12432} provide
 \b*
n^\top F_\sigma (x,\gamma\sigma(x))
 &=&
 \lim_{\eps\rightarrow 0}  \frac{{n^\top  \hat\Delta^\eps_{t}}}{\eps}
 \;\;=\;\;
 \lim_{\eps\rightarrow 0}  \frac{{n^\top \Delta^\eps_{t}}}{\eps}
 \;\;\le\;\;
 0 \;,
 \e*
which concludes the proof.\ep

 \begin{Remark}\label{rm hyperplane1}{\rm 
 Observe that the same line of arguments indicates that \reff{EqViabilityHalfspace} is satisfied for a given vector $n\in \partial \bar B(0,1)$ as soon as the PDE $\L_\sigma u=0$ is first order-viable for an hyperplane $\partial H$ with outward normal vector $n$. One simply needs to work with conditions of the form $n^\top . =0$ instead of $n^\top . \le0$ in the above proof. \\
Moreover, observe that $\partial H$ rewrites $H\cap H'$ with $H'$ a half-space with outward  normal vector $-n$, and  \reff{EqViabilityHalfspace} is automatically satisfied for $-n$ as soon as it is valid for $n$. Therefore, Proposition \ref{PropCondViabSuf}  below indicates that the PDE $\L_\sigma u=0$ is first order-viable for both half spaces $H$ and $H'$ whenever  \reff{EqViabilityHalfspace} holds for $n$. Thus this condition is also necessary and sufficient in order to ensure that the PDE $\Lc_\sigma u=0$ is first order viable for any  hyperplane with outward normal vector $n$.
 }
 \end{Remark}
\begin{Proposition}\label{PropCondViabSuf} 
Suppose that Condition \reff{EqViabilityHalfspace} holds for the vector $n\in\partial \bar B(0,1)$. Then, under \HYP{\sigma} {and $\HYP{_G}$}, the PDE $\Lc_\sigma u=0$ is first order viable for any half-space $H$ with outward normal vector $n$.
\end{Proposition}

 \proof Without loss of generality, we assume that $0 \in \partial H$, recalling Lemma \ref{lem-viab-homothetie}. \\
Let $h$ be a function in $C_b^1(\R^d,\R)$ with gradient valued in $H$. 
    Let $X$ and $(\Delta,\Lambda)$ denote the respective solutions of the SDE \reff{dynamicsRN} and the BSDE \reff{BSDE} associated to any fixed starting point $(t,x)$ in $[0,T]\times \R^d$ and terminal condition $h(X_T)$. We intend to prove that $\Delta$ is valued in $H$ on $[t,T]$. The first order viability of $\Lc_\sigma u=0$ for $H$ is then a direct consequence of Corollary \ref{cor viab BSDE}. 

Using Proposition \ref{pr representation}, we have, for $t\le s \le T$ 
\begin{align} \label{eq temp cond suff}
 n^\top \Delta_s
  &= 
 n^\top \Delta_T+  \int_s^T n^\top  F_\sigma (X_u,\Gamma_u \sigma(X_u))
    du 
  - \int_s^T n^\top \Gamma_u \sigma(X_u) dW_u  \;, \end{align}
 where $\Gamma$ is valued in $S_d$.
 Recalling  the proof of Lemma \ref{le cond hs}, we write $\Gamma_u := \sum_{1\le k \le \ell \le d} \gamma^{k\ell}_u \epsilon_{k \ell}$ and observe that $n^\top \Gamma_u   = \sum_{\ell = 1}^d \gamma^{1\ell}_u n^\top \epsilon_{1 \ell} = \gamma^{1.}_uP^\top$.

Moreover, since $ \Delta^{h}_T=\nabla h(X_T)\in H$, we get from \eqref{eq temp cond suff} that
\begin{align}\label{temptemptemp}
n^\top \Delta_s
  &\le
   \int_s^T \sum_{1\le k \le \ell \le d}  \gamma^{k\ell}_u  \partial_x[ \bar n_\ell^\top \sigma  \sigma^\top \bar n_k ](X_u) n
    du 
  - \int_s^T \gamma^{1.}_u P^\top \sigma(X_u)   dW_u \;, 
 \end{align}
for all $s\in[t, T]$, recalling \eqref{eq reec F}.\\
Using Condition \eqref{EqViabilityHalfspace} in its equivalent form \eqref{eq tru cond hs}, we obtain
\begin{align}\label{eq cond suff temp 2}
 n^\top \Delta_s
  &\le
   \int_s^T \sum_{\ell = 1}^d  \gamma^{1\ell}_u \partial_x[ \bar n_\ell^\top \sigma  \sigma^\top n ](X_u) n
    du 
  - \int_s^T \gamma^{1.}_u P^\top \sigma(X_u)    dW_u \;, 
 \end{align}
for $s\in [t , T]$. 
Defining the vector valued process $\theta_u$ s.t. 
$\theta^\ell_u = \partial_x[ \bar n_\ell^\top \sigma  \sigma^\top n ](X_u) n$, $1 \le \ell \le d$, we compute
 \begin{align}\label{eq cond suff temp 3}
 n^\top \Delta_s
  &\le
  - \int_s^T \gamma^{1.}_u P^\top \sigma(X_u)    (dW_u - [P^\top \sigma(X_u)]^{-1} \theta_u \, d u) \;, \quad t\le s \le T\;.\quad \qquad
 \end{align}
Using the definition of $P$, we notice that 
 \b*
 [P^\top \sigma(X_.)]^{-1} \theta 
 &=& \sigma^{-1}(X_.)\Big(\sum_{1\leq i,j,k\leq d} \partial_k[\sigma\sigma^\top]^{i,j}(X_.)n^i n^k  \big(\sum_{\ell} P^{m\ell} \bar n_\ell^j\big) \Big)_{1\leq m \leq d}
 \\ 
 & = & 
 \sigma^{-1}(X_.)\Big(\sum_{1\leq i,k\leq d} \partial_k[\sigma\sigma^\top]^{i,j}(X_.)n^i n^k \Big)_{1\leq j \leq d}
 \e*
 Therefore,  we can apply Girsanov Theorem under \HYP{_G}, and we know that there exists an  equivalent probability measure $\P^{\theta}$, under which the process $W-\int [P^\top \sigma(X_u)]^{-1} \theta_u du $ is a Brownian motion. Taking the expectation under this new probability measure in \eqref{eq cond suff temp 3}, we obtain
\begin{align*}
n^\top  \Delta_s \le 0, \quad t\le s \le T\;,
\end{align*}
which concludes the proof.
\ep

 \begin{Remark}\label{rm stoptime}{\rm Let $H$ be a half-space with unit outward  normal vector $n$ and $\tau$ a stopping time valued in $[0,T]$. 
 Whenever Condition \reff{EqViabilityHalfspace} holds for $n$, we observe that the same arguments as in the above proof show that the first component of the solution of the BSDE \reff{BSDE} lies in $H$ on $[0,\tau]$, if it belongs to $H$ at time $\tau$. 
 }\end{Remark}

\subsection{First order viability for general convex sets} \label{Subsec viab_main}

 As established in Proposition \ref{prop-viab-hyperplan}, the first order viability of a closed convex set  $K$ with non empty interior is characterized by the first order viability of supporting half-spaces $H_y$ tangent to $K$ at points $y\in\tilde{\partial K}$. We shall verify in this section that it is also characterized in terms of first order viability on the largest class of supporting hyper-spaces $H_y$ associated to any $y\in\breve{\partial K}$. More importantly, we derived in Theorem \ref{ThmViabilityHalfspace}   a necessary and sufficient analytical condition ensuring the PDE $\Lc_\sigma u=0$ to be first order viable for a given half-space. Combining these observations provides therefore a similar condition for any closed convex set $K$ with non empty interior. 

\begin{Theorem}\label{ThmViabilityConvex}
Let $\HYP{_K}$ be in force. If the PDE $\Lc_\sigma u=0$ is first order viable for the convex set $K$, then the condition \reff{Condition}, which rewrites equivalently
  \begin{align}\label{EqViabilityConvex}
 n(y)^\top F_{\sigma}(x,\gamma\sigma(x))  &= 0\;, \quad  \text{for all }~ x\in \text{Supp}(\sigma)\;,\;\; y\in \breve{\partial K}\;,\;\; \gamma\in S_d  \quad \mbox{such that } \gamma\, n(y) =0\;,
 \end{align}
is satisfied. Besides, whenever \HYP{\sigma} and \HYP{_G}  hold, the converse is valid. 
\end{Theorem}

\proof The proof is performed in several steps.

\noindent \textbf{\textit{Step 1:}}\textit{  \reff{EqViabilityConvex} implies the first order viability property.}\\
We assume in this step that \reff{EqViabilityConvex} is satisfied and \HYP{\sigma} holds. Then, for any $y\in\breve{\partial K}$, Theorem \ref{ThmViabilityHalfspace} indicates  that the PDE $\Lc_{\sigma}u=0$ is first order viable for the supporting half-space $H_y$. In particular, it is first order viable for any half-space $H_y$ with $y\in\tilde{\partial K}$ and Proposition \ref{prop-viab-hyperplan} implies that it is first order viable for the convex $K$.\\[-2mm]

\noindent \textbf{\textit{Step 2:}}\textit{  The first order viability property implies \reff{EqViabilityConvex}.}\\
Assume that the PDE $\Lc_{\sigma}u=0$ is first order viable for $K$. Proposition \ref{prop-viab-hyperplan} indicates that this is also true for any half-space $H_y$ with $y\in\tilde{\partial K}$ and Theorem \ref{ThmViabilityHalfspace} implies that
  \b*
 n(y)^\top F_{\sigma}(x,\gamma\sigma(x)) &= 0\;, \qquad  \forall x\in\R^d\;,\;\; y\in \tilde{\partial K}\;,\;\; \gamma\in S_d  \quad \mbox{ such that } \;\;\; \gamma\,  n(y) =0\;.
 \e*
 It remains to check that this relation is also valid for a given $y\in \breve{\partial K}$.\\ 
 Let $y$ be in $\breve{\partial K}$. As observed in  the proof of Lemma \ref{LemmaDecompConv}, $y\in \breve{\partial K}$ is the limit of a sequence $(y_p)_{p\in\N}$ of points lying in $\tilde{\partial K}$ s.t. $n(y_p) \rightarrow n(y)$. Fix now some $(x,\gamma)\in\R^d\times S_d$ satisfying $\gamma\, n(y) =0$. Then there exists a sequence $(\gamma_p)$ in $S_d$ such that  $\gamma_p \, n(y_p)=0$ for any $p\geq 1$ and $\gamma = \lim_{p\rightarrow\infty}\gamma_p$. Indeed, since $\gamma\in S_d$, there exists  $O\in M_d$ s.t. $O O^\top=O^\top O=I_d$ and $O^\top \gamma O =D$ where $D$ is a diagonal matrix with $D^{1,1}=0$. The matrix $O$ corresponds to the new basis matrix from the canonical basis  to an orthonormal basis $B'=(n(y),n_2,\ldots,n_d)$. Then consider the basis $B'_p$ obtained by applying the Gram-Schmidt orthonormalization procedure to the basis $(n(y_p)_p,n_2,\ldots,n_d)$ and denote by $O_p$ the new basis matrix for $B'_p$ for $p\geq 1$. Since the Gram-Schmidt orthonormalization procedure is continuous, we get that $\lim_{p\rightarrow\infty}O_p=O$. Then define the sequence $(\gamma_p)_p$ in $S_d$ by
 \b*
 \gamma_p & := & O_p^\top D O_p\;,\quad p\geq 1\;.
 \e*
 From the definition of the basis $B'_p$ and since $D_{1,1} =0$ we have $ \gamma_p \, n(y_p) = 0$ for all $p\geq 1$. Moreover, since $\lim_{p\rightarrow\infty}O_p=O$, we have $\lim_{p\rightarrow\infty}\gamma_p=\gamma$.\\
Now, since $y_p\in \tilde{\partial K}$ and $ \gamma_p \, n(y_p) = 0$, we get 
\b*
 n(y_p)^\top F_\sigma(x,\gamma_p\sigma(x)) & =& 0\;,
\e*  
 for any $p\geq 1$. Letting $p$ go to infinity, we obtain $ n(y)^\top F_\sigma(x,\gamma\sigma(x)) = 0$ and \reff{EqViabilityConvex}  holds also for $(y,x,\gamma)\in \breve{\partial K}\times\R^d\times S_d$ such that $\gamma \,  n(y)=0$.
\\[-2mm]
\ep

  \begin{Remark}\label{rm stoptime2}{\rm 
(i) As observed in Remark \ref{rm stoptime}, Condition \reff{Condition} also ensures that, if the first component $\Delta$  of the solution to the BSDE \reff{BSDE} lies in $H$ at a given stopping time $\tau$ valued in $[0,T]$, $\Delta$ remains in $K$ on $[0,\tau]$. Therefore Condition \reff{Condition} is necessary and sufficient  to ensure that the BSDE  \reff{BSDE} satisfies the first order viability property for $K$ on any random time interval $[0,\tau]$, with $\tau$ stopping time smaller than $T$. 
 
(ii) Consider for example an American option whose exercise payoff is $h(X^{t,x})$ on $[t,T]$. We assume that $h\in C^1_b(\R^d,\R)$ with $\nabla h$ valued in $K$. We denote by $\tau^*$ the optimal stopping time. Under some regularity assumptions, it is known that the {\it Delta} of the option at time $t$ is $\Delta^{t,x}_t$, where the terminal condition in \eqref{BSDE} is now random and given by $\nabla h(X^{t,x}_{\tau^*})\in K$, see e.g. \cite{gob04} Theorem 2.3 and the references therein. One can then apply (i) above to conclude that Theorem \ref{maintheorem} holds true for American Option, under strengthened regularity assumption. }\end{Remark}

%
%

\section{Proof of Theorem \ref{maintheorem}} \label{SectionRegu}

In this Section we prove the main result of this paper, i.e. Theorem \ref{maintheorem}, using the results of Section 4 together with some regularization arguments. We prove each implication separately. 

\subsection{(i) $\implies$ (ii)}
Let $h$ be any  function in $C^1_b(\R^d,\R)$ such that $\nabla h$ is valued in $K$. Our goal is to show that $\nabla u^h$ is valued in $K$ when (i) holds true ($h$ is not necessarily bounded from below). If this is the case, then the PDE $\Lc_\sigma u = 0$ is first-order viable and the statement is a straightforward consequence of Theorem \ref{ThmViabilityConvex}.
\\
We now construct an approximating sequence of functions $ (h_n)_{n\ge0}$ s.t. $h_n$ satisfies 
\HYP{h} and $\nabla u^{h_n}(t,x) \rightarrow \nabla u^h(t,x)$, as $n \rightarrow \infty$.
\\
To this end, we introduce, for $n \ge 0$,
\b*
f_n: z\in\R &  \mapsto &   \int_{-\infty}^z F_{[0,1]}[\1_{[-n,\infty)}](y) d y-\Big(n+{1\over 2}\Big)\;,
\e*
where $F_{[0,1]}$ is the facelift operator on $\R$ associated to the convex set $[0,1]$. 
We then define the sequence $(h_n)_n$ by  $h_n = f_n \circ h$ for all $n\geq 0$. We notice that $f_n$ is lower bounded and $f_n(y)=y$ for all $y\geq -n$, and all $n\geq 0$.
We compute that $\nabla h_n = f'_n( h) \nabla h$  and since $f'_n \in [0,1]$, $\nabla h_n \in [0,\nabla h ] \subset K$. Thus $h_n$ satisfies $\HYP{h}$. Moreover,
we have $h_n(y) \rightarrow h(y)$ and $\nabla h_n(y) \rightarrow \nabla h(y)$, as $n \rightarrow \infty$, for all $y\in \R^d$. Using the representation of Proposition \ref{pr representation} and usual stability arguments for BSDEs (see e.g. Proposition 2.1 in \cite{karpen97}), we obtain that
$\nabla u^{h_n}(t,x) \rightarrow \nabla u^h(t,x)$, as $n \rightarrow \infty$. Under (i), we have that $\nabla u^{h_n}$ takes its values in $K$ and so does $ \nabla u^{h}$.  \eproof

\subsection{(ii) $\implies$ (i)}
\label{subse ii=>i}

\noindent \textbf{\textit{Step 1:}}\textit{ Replication strategy for $\Lift{h}(X_T^{t,x})$.} \\
Let fix $(t,x)\in[0,T)\times\R^d$.  Under \HYP{h}, $h$ is bounded from below. Hence,  Lemma \ref{le basic prop} (v) ensures that $\Lift{h}(X^{t,x}_T)$ is also bounded from below. Using the martingale representation Theorem, we have
\begin{align*}
u^{\Lift{h}}(t,x) = \Lift{h}(X^{t,x}_T) - \int_t^T Z^{t,x}_s d W_s,
\end{align*}
which allows us to define the replicating financial strategy $(\Delta_s)_{s\in[t,T]}$ for $\Lift{h}(X^{t,x}_T)$ by  
\b*
\Delta_s^\top & := & Z^{t,x}_s\sigma^{-1}(X^{t,x}_s)\;,\quad s\in [t,T]\;. 
\e* 
Since $Z^{t,x}\in\Hc^2[t,T]$ and $\sigma^{-1}(X^{t,x})$ is a continuous process, this strategy is obviously admissible.

\noindent \textbf{\textit{Step 2:}}\textit{ Viability of the replication strategy of $\Lift{h}$.} \\
 We now  prove that the replicating strategy  $\Delta$ is admissible.
The main difficulty relies here in the lack of regularity of the payoff function under assumption \HYP{h}. We therefore use an approximation argument and proceed in two substeps. 

\noindent \textbf{\textit{Substep 2.a:}}\textit{ Regularization of $h$.} \\ Under \HYP{h}, we consider the Lipschitz-regularization $(h_n)$ of $h$ given in Lemma \ref{le reg 1}. We introduce the sequence $({}^nZ^{t,x})$ given by
\b*
u^{\Lift{h_n}}(t,x) & = & \Lift{h_n}(X^{t,x}_T) - \int_t^T {}^nZ^{t,x}_s d W_s\;, \qquad n\ge 1\;.
\e*
Using \HYP{h}, Lemma  \ref{le reg 1} and the dominated convergence theorem, we easily obtain
\begin{align*}
\lim_{n \rightarrow\infty}\esp{|\Lift{h}(X^{t,x}_T) - \Lift{h_n}(X^{t,x}_T)|^2} = 0
\end{align*} 
and thus ${}^nZ^{t,x} \rightarrow Z^{t,x}$ in $\Hc^2[t,T]$.
\\
Defining the process 
$
{}^n\Delta^\top := {}^nZ^{t,x}\sigma^{-1}(X^{t,x})
$
we directly deduce that up to a subsequence  
\be \label{eq cv strat 1}
{}^n\Delta &  \rightarrow  & \Delta\;,\quad  \P\otimes \lambda ~ a.e. \text{ on } \Omega\times [t,T]
\;.
\ee

\noindent \textbf{\textit{Substep 2.b:}}\textit{ Regularization of $\Lift{h_n}$.}
\\
Fix $n\in\N$ and let $\varphi$ be  a compactly supported smooth probability density function on $\R^d$. 
\\
We define the sequences of function $(\varphi_k)_{k\ge 1}$ and $(F_{n,k})_{k\ge 1}$ from $\R^d$ to $\R$ by
$
\varphi_k: x\mapsto k^d \varphi(k x)\;
$
and 
\begin{align*}
F_{n,k} : x \mapsto \varphi_k * \Lift{h_n}(x)~=~\int_{\R^d}\varphi_k(x+y) \Lift{h_n}(x)dy\,, 
\end{align*}
 for any $k\geq 1$.
 Let us introduce ${}^{k,n}Z$ given by the martingale representation Theorem
\begin{align*}
u^{F_{n,k}}(t,x) =  \varphi_k * \Lift{h_n}(X_T^{t,x}) - \int_t^T {}^{k,n}Z^{t,x}_s d W_s \;, \qquad n,k\ge 1 \;.
\end{align*}
As in the previous step, we define the sequence of processes ${}^{k,n}\Delta^\top := {}^{k,n}Z^{t,x} \sigma^{-1}(X^{t,x})$.
We observe that, up to a subsequence,  ${}^{k,n}\Delta \rightarrow {}^{n}\Delta$ a.e. on $\Omega\times[t,T]$ as $k$ goes to $\infty$. Besides, Theorem 3.1 in \cite{majzha02} directly implies that 
\begin{align}
{}^{k,n}\Delta_s = \nabla u^{F_{n,k}}(s ,X^{t,x}_s) \;, \qquad t\le s \le T   \;, \qquad n,k\ge 1 \;. \label{eq rep strat}
\end{align}

\noindent  Now, since $\Lift{h_n}$ is a bounded Lipschitz function, 
combining Rademacher Theorem with
Lemma \ref{le pde Lift} (i) we have 
$
\mathcal{C}_K \big( \partial_x \Lift{h_n}(x) \big)   \ge   0 
$,
for  almost every $x\in \R^d$, 
which means
\b*
\nabla \Lift{h_n} &  \in & K \;, \quad \text{ a.e. on }~\R^d \;,
\e*
We also observe that, since $\Lift{h_n}$ is Lipschitz cotinuous, we have from the dominated convergence theorem
\b*
\partial_j  F_{n,k}  & = &  \varphi_k * \partial_j \nabla \Lift{h_n},  \qquad 1\le j \le d\;.
\e*

\noindent Now, since $K$ is closed and convex, we obtain
\be
\nabla F_{n,k}  & \in & K \;, \qquad n,k\ge 1 \;.  \label{eq temp term cond}
\ee
Applying Theorem  \ref{ThmViabilityConvex}, we deduce that
$(\nabla u^{F_{n,k}}(t,.) )_{n,k}$ and thus $({}^{n,k}\Delta)_{n,k}$ are valued in $K$, recalling \eqref{eq rep strat}.

\noindent \textbf{\textit{Substep 2.c:}}\textit{ Viability of the replicating strategy of $F_K[h]$.}
\\
For any $n\ge 1$, since ${}^{k,n}\Delta \rightarrow {}^{n}\Delta$ a.e. as $k$ goes to infinity, 
 the closeness of $K$ implies that  ${}^{n}\Delta$ is valued in $K$ a.e. 
\\
A similar argument yields that $\Delta$ is also valued in $K$, recalling Step 1. 
Since $\Delta$ is an admissible strategy, we conclude that $\Delta \in \mathcal{A}^K_{t,x}$.

\noindent \textbf{\textit{Step 3:}}\textit{ Identification of the super replicating price of $h$ and the replicating price of $\Lift{h}$.} \\
 Substep 2.c yields that $u^{\Lift{h}}(t,x)$ dominates the super-replication price $v^h_K(t,x)$ of $h(X^{t,x}_T)$, recalling Definition \ref{de super-replication}. The proof is concluded using Corollary \ref{co carac utile}.
\eproof

%
%

\section{Appendix}

\subsection{Facelift properties}
\label{subse facelift}

The first lemma collects some useful properties of the facelift transform. Lemma \ref{le reg 1} is an approximation result and Lemma \ref{le pde Lift} is a (minimal) PDE characterisation of the facelift.

\begin{Lemma} \label{le basic prop}

(i) If h is lower semi-continuous, then $\Lift{h}$ is also l.s.c..

(ii) If $h(x) \ge g(x)$ for all $x$, then $\Lift{h}(x) \ge \Lift{g}(x)$, for all $x\in\R^d$.

(iii) If $0 \in K$ and $h(.)=c$ with $c$ a given constant then $\Lift{h}=h$.

(iv) $\Lift{h\vee g} = \Lift{h} \vee \Lift{g}$.

(v) $\Lift{h} \ge h$ and $\Lift{h} = \Lift{\Lift{h}}$.
\end{Lemma}

\proof Property (i) holds true since $\Lift{h}$ is  the point wise supremum of l.s.c. functions. Properties (ii)-(iii)-(iv) are obvious consequences of the Definition \ref{defFacelift} of the facelift transform. Property (v) follows from the fact that
 $0 \in \tilde K$ and the following computation
 \begin{align*}
\Lift{\Lift{h}}(x) &= \sup_{y_2 \in \R^d} \Lift{h}(x+y_2) - \delta_K(y_2)
\\
&= \sup_{y_1,y_2 \in \R^d} h(x+y_2+y_1)-  \delta_K(y_1) - \delta_K(y_2)
\;\le\; \Lift{h}(x)\;,
\end{align*}
for any $x\in\R^d$.
\eproof

\begin{Lemma} \label{le reg 1}
Assume that $h$ is lower semi-continuous and bounded from below by $-m_h$, for some $m_h \ge 0$.
\\
Then, there  exists an increasing sequence $(h_n)_{n \ge 1}$ of
bounded Lipschitz function, uniformly bounded from below by $-m_h$ converging to $h$ and such that
$\Lift{h_n} \uparrow \Lift{h}$. 
\end{Lemma}
\proof  We define the sequence of functions $(g_n)_n$ by
\b*
g_n(x) &  = &  \inf_{y \in \R^d}\big\{h(y) + n|x-y|\big\} \;,\quad x\in \R^d\;, 
\e*
for $n\geq 1$.
It is clear that the sequence $(g_n)_n$ is nondecreasing, that 
 $-m_h \le g_n \le h$ 
 and $g_n$ is $n$-Lipschitz continuous for all $n \geq 1$.

We now prove that $(g_n)_n$ converges pointwise to $h$. Fix some $x\in\R^d$.
Since $h$ is l.s.c and bounded from below there exists a sequence $(x_n)_n$ in $\R^d$
such that
\be\label{def x n }
g_n(x) &  = &  h(x_n) + n |x-x_n|\;, \qquad n\ge 1\;.
\ee
Since $h$ is bounded from below by $-m_h$, we deduce
\b*
n |x-x_n| & \le &  h(x) - h(x_n) ~ \le ~ h(x) + m_h\;,  \qquad n\ge 1\;,
\e*
 so that   $\lim_{n\rightarrow\infty}x_n = x$. Together with  \reff{def x n } and  the lower semi continuity of $h$, this yields
\b*
\lim_{n\rightarrow\infty} g_n(x) &  \ge &  \liminf_{n \rightarrow \infty} h(x_n) ~ \ge ~  h(x)\;.
\e*
Thus,  $g_n(x) \uparrow h(x)$ as $n\uparrow \infty$, for all $x \in \R^d$.
\\
 Define now the sequence of functions $(h_n)_n$ by 
\b*
h_n (x)  & := &  g_n(x) \wedge n\;,\quad x\in\R^d\;, \;,  \qquad n\ge 1\;.
\e*
Since $g_n$ is Lipschitz continuous and bounded from below, $h_n$ is bounded and Lipschitz continuous, for all $n\geq 1$. Moreover, since $(g_n)_n$ is nondecreasing and converges pointwise to $h$, we also get that   $(h_n)_n$ is nondecreasing and converges pointwise to $h$.
%

It remains to prove the convergence of  $\Lift{h_n}$ to $ \Lift{h}$.  For any $x\in\R^d$, we simply observe that 
\b*
\Lift{h}(x) &= & \sup_{u \in \tilde{K}} h(x+u)-\delta_K(u) 
			\;=  \; \sup_{n\ge 1, u \in \tilde{K}} h_n(x+u)-\delta_K(u) 
			\;= \;  \lim_{n\rightarrow\infty} \uparrow  \Lift{h_n}(x)\;.
\e* 
\eproof

\begin{Lemma} \label{le pde Lift} Let $h$ be a lower semi-continuous function from $\R^d$ to $\R$.

\noindent (i) Assume that $\Lift{h}$ is locally bounded, then $\Lift{h}$ is a
viscosity super-solution of
\be
 \min \set{ \mathcal{C}_K(\partial_x u), u-h} & = & 0  \qquad \mbox{on}\quad  \R^d\;.  \label{eq edp lift}
\ee
(ii) Let $v$ be a differentiable super-solution of \eqref{eq edp lift}, then
\b*
v(x)  & \ge &  F_K[v](x) ~ \ge ~  F_K[h](x) \;, \qquad x\in\R^d\;.
\e*
In particular, if $h$ is differentiable and $\nabla h \in K$, then $F_K[h] = h$.
\end{Lemma}

\proof \noindent \textbf{\textit{Step 1:}} Proof of (i). \\
We first recall that, since $h$ is l.s.c continuous, $\Lift{h}$ is also l.s.c, see Lemma \ref{le basic prop}.  Let $\bar x \in \R^d$ and $\phi\in C^1(\R^d,\R)$   a test function such
that
\be\label{test func super sol}
0 & = & \Lift{h}(\bar x) - \phi(\bar x) ~= ~  \text{(strict)}\min_{x \in \R^d} (\Lift{h}-\phi)(x) \;.
\ee
 Observe that Lemma \reff {le basic prop}  (v) implies $ \Lift{\Lift{h}}=\Lift{h}$, so that
\b*
\Lift{h}(\bar x) &  \ge &  \Lift{h}(\bar x + y) - \delta_K(y) \;,   \qquad y\in\R^d\;. 
\e*
Using \reff{test func super sol}, we deduce
\b*
\phi(\bar x)  & \ge &  \phi(\bar x + y) - \delta_K(y) \;,   \qquad y\in\R^d\;. 
\e*
In particular, for $y = \eps \zeta$ where $\eps>0$ and $\zeta\in \tilde K$ with $|\zeta|=1$, we obtain
\b*
\frac{\phi(\bar x) - \phi(\bar x + \epsilon\zeta) }{\epsilon} &  \le &  - \delta_K(\zeta) \;,   \qquad y\in\tilde{K}\;.
\e*
Letting $\eps$ goes to $0$ yields $ \delta_K (\zeta) - \partial_x \phi (\bar
x ) \zeta \ge 0$. Since $\zeta$ is arbitrarily chosen in $\tilde K$, this yields $\Cc_K(\partial_x\phi)(\bar x)\geq 0$.

\vspace{2mm}

\noindent \textbf{\textit{Step 2:}} Proof of (ii).\\ 
Let $v$ be a differentiable supersolution of \eqref{eq edp lift}. We then have 
\b*
\Supp(y) - \partial_xv(x+ty)'y & \ge & 0 \;, \qquad (t,x,y)\in[0,T]\times\R^d\times\tilde{K}\;.
\e*
 Fix now $\bar x \in \R^d$. We get from the previous inequality
\b*
\int_0^T\Big(\Supp(y) - {\partial v(\bar x + t y)\over \partial t} \Big) d t  & \ge & 0 \;, \qquad y\in\R^d\;.
\e*
Therefore, we compute
\b*
v(\bar x) & \ge & v(\bar x+y) -\Supp(y)  \;, \qquad y\in\R^d\;.
\e*
 Taking the supremum over $y$, we obtain $v (\bar x)\ge \Lift{h} (\bar x)$.


Suppose now that $h$ is differentiable and  $[\partial_x h]^\top \in K$. Since we already know that $\Lift{h} \ge h$, we conclude $F_K[h] = h$.
\eproof

\subsection{Proof of Corollary \ref{co carac utile}}\label{Appendix2}

For sake of clarity, let us define $\tilde{v}^h_K$ by
\b*
\tilde{v}^h_K(t,x) & := & v^h_K(t,x) \quad\text{ for } (t,x)\in[0,T)\times\R^d\;, \\
\tilde{v}^h_K(T,x)  & :=  & v^h_K(T-,x)\quad\text{ for } x \in \R^d\;. 
\e*
Recall also the definition of $u^{\Lift{h}}$ and $(u^{\Lift{h_n}})_n$ in the proof of Theorem 3.1, Section \ref{subse ii=>i}, Step 1 and Substep 2.a and the fact that
\b*
\Lift{h}(x) ~ = ~\lim_{n\infty}  \uparrow \Lift{h_n}(x) &\text{ and } & u^{\Lift{h}}(t,x) ~=~ \lim_{n\infty} \uparrow u^{\Lift{h_n}}(t,x) \;,
\e*
for $(t,x) \in [0,T]\times \R^d$.
\\
From the left-hand equality of the above statement and Proposition \ref{pr bruno}, we deduce that $\tilde{v}^h_K$ is a viscosity super-solution of 
\begin{align} \label{eq linear pde}
\left \{
\begin{array}{rccl}
- \mathcal{L}_{\sigma} u(t,x) & = & 0  & \text{ for } ~(t, x) \in [0,T)\times \R^d\;,
\\
u(T,x) &=& \Lift{h_n}(x)   &  \text{ for }~ x \in \R^d\;.
\end{array}
\right. 
\end{align}
Since $\Lift{h_n}$ is Lipschitz continuous, it is also well known (see e.g.  \cite{parpen92}) that $u^{\Lift{h_n}}$ is a viscosity solution of \eqref{eq linear pde}, for any $n\ge1$.
\\
The PDE \eqref{eq linear pde} satisfies the assumptions of Theorem 4.4.5 in \cite{pham09}, which provides a strong comparison theorem for viscosity solutions with polynomial growth. Since the functions $u^{\Lift{h_n}}$ and $\tilde{v}^h_K$ have linear growth, this yields 
\b*
\tilde{v}^h_K(t,x) & \ge & u^{\Lift{h_n}}(t,x) \;, \qquad (t,x) \in[0,T]\times \R^d\;,
\e*
for any $n\ge1$. The proof is  concluded letting $n$ go to infinity.
\eproof

\subsection{Proof of Lemma \ref{lemma approx Delta}}\label{Appendix3}

We fix $(t,x)\in[0,T]\times\R^d$. We first notice that, since $\gamma\in S_d$, the terminal condition $\gamma(X^{t,x}_T-x)$ can be written under the form $\gamma(X^{t,x}_T-x)= \partial_x \bar h(X^{t,x}_T)$ with $\bar h$ defined by 
\b*
\bar h(x') & = & {1\over 2} (x'-x)^\top \gamma (x'-x) \;,\qquad x'\in\R^d\;.
\e*
However, we cannot directly conclude from the $H$-first order viability of $\Lc_\sigma u=0$ that $\Delta^{t,x}$ belongs to $H$, since the terminal payoff function  $h$ does not belong to $C_b^1(\R^d,\R)$. We therefore construct a sequence $(h_p)_p$ valued in $C^1_b(\R^d,\R)$  approximating $h$.   

 Since $\gamma\, n=0$,  we can write $\gamma = \gamma^++\gamma^-$ with $\gamma^+$ and $\gamma^-$ two elements of $S_d$ which are respectively non-negative and non-positive and satisfy $\gamma^+\, n=\gamma^-\, n==0$. \\
 For all $p\geq 1$, define the function $h_p$ by
\b*
 h_p(x') & = &  h_p^+(x')+h_p^-(x')\;,\quad x'\in\R^d\;,  
\e*
where the function $h^+_p$ is defined by 
\begin{equation*}
 h_p^+(x') ~ = ~ 
\left\{
\begin{array}{ccc}
p\Big( |\sqrt{\gamma ^+}(x'-x)|-{p\over 2} \Big) & \mbox{  if } & |\sqrt{\gamma^+}(x'-x)| \geq p\;, \\
{1\over 2} (x'-x)^\top \gamma^+ (x'-x)  &\mbox{  if } &  |\sqrt{\gamma^+}(x'-x)| \leq p \;,
\end{array}\right.
\end{equation*}
and $h_p^-$ is defined by 
\begin{equation*}
 h_p^-(x') ~ = ~ 
\left\{
\begin{array}{ccc}
-p\Big( |\sqrt{-\gamma ^-}(x'-x)|-{p\over 2} \Big) & \mbox{  if } & |\sqrt{-\gamma^-}(x'-x)| \geq p\;, \\
{1\over 2} (x'-x)^\top \gamma^- (x'-x)  &\mbox{  if } &  |\sqrt{-\gamma^-}(x'-x)| \leq p \;,
\end{array}\right.
\end{equation*}
for all $x'\in\R^d$. We then easily check that
\begin{equation*}
\partial_x h_p^\pm(x')^\top  ~ = ~ 
\left\{
\begin{array}{ccc}
  p{\gamma^\pm (x'-x) \over |\sqrt{\pm\gamma^\pm} (x'-x)| }& \mbox{  if } & |\sqrt{\pm\gamma^\pm}(x'-x)| \geq p\;, \\
\gamma^\pm (x'-x)   &\mbox{  if } &  |\sqrt{\pm\gamma^\pm}(x'-x)| \leq p \;.
\end{array}\right.
\end{equation*}
for all $x'\in\R^d$. Therefore we get from the dominated convergence theorem that
\be\label{approx cd terminale}
\E\Big[\Big|\partial_x  h_p(X^{t,x}_T)-\partial_x  \bar h(X^{t,x}_T)\Big|^2\Big] & \longrightarrow &  0\;,\quad \mbox{ as }~~p\rightarrow \infty\;.
\ee
 Observe that $h_p\in C^1_b(\R^d,\R)$ and $\partial_x  h_p$ is valued in $H$, for all $p\geq 1$. 
Since the PDE $\Lc_\sigma u=0$ is first order viable for $H$, we deduce from Proposition \ref{pr representation} that
\be\label{approx viab}
\Delta_s^{p} & \in & H\;,\qquad  t\le s \le T\;, \qquad p\ge 1\;,
\ee
where $(\Delta^p,\Lambda^p)$ is the solution on $[t,T]$ of the BSDE \reff{BSDE} associated to the terminal condition $\partial_x h_p(X^{t,x}_T)^\top$.

We get from  \reff{approx cd terminale} and classical estimates on BSDEs that 
\b*
\E\Big[\sup_{t\leq s \leq T}\Big|  \Delta^{t,x}_s- \Delta_s^p\Big|^2\Big] & \longrightarrow &  0\;,\quad \mbox{ as }~~p\rightarrow \infty\;.
\e*
Since $H$ is closed, \reff{approx viab} together with  the previous convergence imply that 
$\Delta^{t,x}$ is valued in $H$.

\eproof

\newpage

\end{document}